\documentclass[journal=jacsat,manuscript=article,layout=twocolumn]{achemso}
\usepackage[utf8]{inputenc}
\usepackage{svg}
\usepackage{amsmath,bm}
\usepackage{mathrsfs}
\usepackage{amsfonts}
\usepackage{float}
\usepackage{setspace} 
\usepackage{graphicx}
\usepackage{epstopdf}
\usepackage{dcolumn}
\usepackage{amsmath}
\usepackage{epsfig}
\usepackage{indentfirst}
\usepackage{psfrag}
\usepackage{subfigure}
\usepackage{amssymb}
\usepackage{color}
\usepackage{physics}
\usepackage{natbib}
\usepackage{wrapfig}
\usepackage{siunitx}

\usepackage{ulem}[normalem] 

\def\ttat{\textbf{TTAT}}

\def\t2tat{\textbf{T2TAT}}
\def\ef{E$_{\tiny{\mbox{F}}}$} 

\newcommand{\onlinecite}[1]{\hspace{-1 ex} \nocite{#1}\citenum{#1}}
 
\usepackage{xr}

\makeatletter
\newcommand*{\addFileDependency}[1]{
  \typeout{(#1)}
  \@addtofilelist{#1}
  \IfFileExists{#1}{}{\typeout{No file #1.}}
}
\makeatother
\newcommand*{\myexternaldocument}[1]{%
    \externaldocument{#1}%
    \addFileDependency{#1.tex}
    \addFileDependency{#1.aux}%
}
\myexternaldocument{Supplementary}
\graphicspath{ {Fig/} {Figsi/} }  

\usepackage[backref=none,bookmarksnumbered=true,bookmarks=true,bookmarksopen=true,colorlinks=true,citecolor=blue,linkcolor=blue,anchorcolor=green,urlcolor=blue,unicode=false]{hyperref}

\let\oldmaketitle\maketitle
\let\maketitle\relax

\title{On-surface Synthesis of \\ a Ferromagnetic Molecular Spin Trimer}

 \author{Alessio Vegliante}
\affiliation{CIC nanoGUNE-BRTA, 20018 Donostia-San Sebasti\'an, Spain}
\altaffiliation{Contributed equally to the work}

\author{Manuel Vilas-Varela}
\affiliation{Centro Singular de Investigaci\'on en Qu\'imica Biol\'oxica e Materiais Moleculares (CiQUS) and Departamento de Qu\'imica Org\'anica, Universidade de Santiago de Compostela 15782-Santiago de Compostela (Spain)}
\altaffiliation{Contributed equally to the work}

\author{Ricardo Ortiz}
\affiliation{Donostia International Physics Center (DIPC), 20018 Donostia-San Sebastián, Spain}

\author{Francisco Romero Lara}
\affiliation{CIC nanoGUNE-BRTA, 20018 Donostia-San Sebasti\'an, Spain}

\author{Manish Kumar}
\affiliation{Institute of Physics, Czech Academy of Sciences, Prague, Czech Republic}

\author{Lucía Gómez-Rodrigo}
\alsoaffiliation{Centro Singular de Investigaci\'on en Qu\'imica Biol\'oxica e Materiais Moleculares (CiQUS) and Departamento de Qu\'imica Org\'anica, Universidade de Santiago de Compostela 15782-Santiago de Compostela (Spain)}

\author{Fabian Schulz}
\affiliation{CIC nanoGUNE-BRTA, 20018 Donostia-San Sebasti\'an, Spain}

\author{Diego Soler}
\affiliation{Institute of Physics, Czech Academy of Sciences, Prague, Czech Republic}

\author{Hassan Ahmoum}
\affiliation{CIC nanoGUNE-BRTA, 20018 Donostia-San Sebasti\'an, Spain}

\author{Emilio Artacho}
\affiliation{CIC nanoGUNE-BRTA, 20018 Donostia-San Sebasti\'an, Spain}
\alsoaffiliation{Ikerbasque, Basque Foundation for Science, 48013 Bilbao, Spain}
\alsoaffiliation{Donostia International Physics Center (DIPC), 20018 Donostia-San Sebastián, Spain}
\alsoaffiliation{Theory of Condensed Matter, Cavendish Laboratory, University of Cambridge, J. J. Thomson Avenue, Cambridge CB3 0HE, United Kingdom}

\author{Thomas Frederiksen}
\affiliation{Donostia International Physics Center (DIPC), 20018 Donostia-San Sebastián, Spain}
\alsoaffiliation{Ikerbasque, Basque Foundation for Science, 48013 Bilbao, Spain}

\author{Pavel Jelínek}
\affiliation{Institute of Physics, Czech Academy of Sciences, Prague, Czech Republic}
\alsoaffiliation{Czech Advanced Technology and Research Institute (CATRIN), Palacký University Olomouc, 779 00 Olomouc, Czech Republic}
 
\author{Jose Ignacio Pascual}
\affiliation{CIC nanoGUNE-BRTA, 20018 Donostia-San Sebasti\'an, Spain}
\alsoaffiliation{Ikerbasque, Basque Foundation for Science, 48013 Bilbao, Spain}
 \email{ji.pascual@nanogune.eu}

\author{Diego Peña}
\affiliation{Centro Singular de Investigaci\'on en Qu\'imica Biol\'oxica e Materiais Moleculares (CiQUS) and Departamento de Qu\'imica Org\'anica, Universidade de Santiago de Compostela 15782-Santiago de Compostela (Spain)}
\alsoaffiliation{Oportunius, Galician Innovation Agency (GAIN)
15702 Santiago de Compostela, Spain}
\email{diego.pena@usc.es}

\begin{document}
\date{\today}

\twocolumn[
\begin{@twocolumnfalse}
\oldmaketitle
\begin{abstract}
\setstretch{0.9}
\noindent\hrulefill
\vspace{2mm}

Triangulenes are prototypical examples of open-shell nanographenes. Their magnetic properties, arising from the presence of unpaired $\pi$ electrons, can be extensively tuned by modifying their size and shape or by introducing heteroatoms. Different triangulene derivatives have been designed and synthesized in recent years, thanks to the development of on-surface synthesis strategies. Triangulene-based nanostructures with polyradical character, hosting several interacting spin units, can be challenging to fabricate but are particularly interesting for potential applications in carbon-based spintronics. Here, we combine pristine and N-doped triangulenes into a more complex nanographene, \textbf{TTAT}, predicted to possess three unpaired $\pi$ electrons delocalized along the zigzag periphery. We generate the molecule on an Au(111) surface and detect direct fingerprints of multi-radical coupling and high-spin state using scanning tunneling microscopy and spectroscopy. With the support of theoretical calculations, we show that its three radical units are localized at distinct parts of the molecule and couple via symmetric ferromagnetic interactions, which result in a $S=3/2$ ground state, thus demonstrating the realization of a molecular ferromagnetic Heisenberg-like spin trimer. 

\noindent\hrulefill
 \end{abstract}
\end{@twocolumnfalse}
]

\setstretch{0.9}

\textbf{Introduction}:   Recent advances in molecular nanoscience have shown that small graphene flakes with atomically customized shapes can exhibit $\pi$-paramagnetism associated with radical states in open-shell structures. This unique form of magnetism exhibits distinctive characteristics such as spin delocalization and large spin exchange interactions \cite{Li2019, Mishra2020}. Owing to the weak spin-orbit coupling in carbon compounds, spin-hosting nanographenes (NGs) are anticipated to exhibit long spin-coherence times, making them promising candidates for applications in quantum computing \cite{Gaita2019}. The spin ground state of a magnetic nanographene is determined by its atomic-scale structure and composition  \cite{Yazyev2010,deOteyza2022}. Therefore, atomically precise on-surface synthesis (OSS) techniques \cite{Clair2019}, in combination with the solution synthesis of organic precursors, offer a unique opportunity to engineer novel magnetic states in two dimensions through the creation of unpaired electrons at radical sites.

Graphene flakes with triangular shapes are a paradigmatic platform for hosting interacting quantum spins. The  Ovchinnikov's rule \cite{Ovchinnikov1978} for alternant conjugated lattices predicts an intrinsic spin imbalance. The net spin of triangulene molecules, for example, increases with the flake's size and can also be modified by heteroatoms substitution \cite{Pavlicek2017, Turco2023, Mishra2019, Su2019, Mishra2021b, Wang2022, Vilas-Varela2023, Lawrence2023}. 
The high-spin states of triangulenes are exceptionally robust because their singly-occupied orbitals (or zero-energy states) live in the same carbon sublattice, thus having a large spatial wavefunction overlap. Consequently, Hund's exchange coupling and spin excitation energies generally amount to a significant fraction of an electronvolt. While systems with considerable energy gaps between the ground and excited spin states can be attractive for some applications related to classical magnetism, this can be a drawback for applications utilizing the full spin  spectrum of the nanographene.

Nanographenes with weakly interacting spins have been successfully synthesized on surfaces by covalently-bonding triangulene building blocks through their vertices \cite{Mishra2020b,Zheng2020,Mishra2021b,Hieulle2021,Cheng2022,Du2023,Turco2024}. This strategy maintains the triangulene integrity because the zero-energy modes have a low density of states over these connecting sites.  As a result, collective spin states emerge in polyradical chains and rings from antiferromagnetic interactions between the monomer units. However, there is scarce direct evidence of ferromagnetic exchange interactions between weakly coupled radical states that would allow us to investigate the full spin excitation spectrum. An alternative strategy rarely explored is connecting the nanographenes through their zigzag edges. While this method modifies the electronic and magnetic configuration of the original structure \cite{Calupitan2023}, it can build customized flakes with interacting localized radicals \cite{Song2024}.

\begin{figure*} [h!]
   \includegraphics[width=\textwidth]{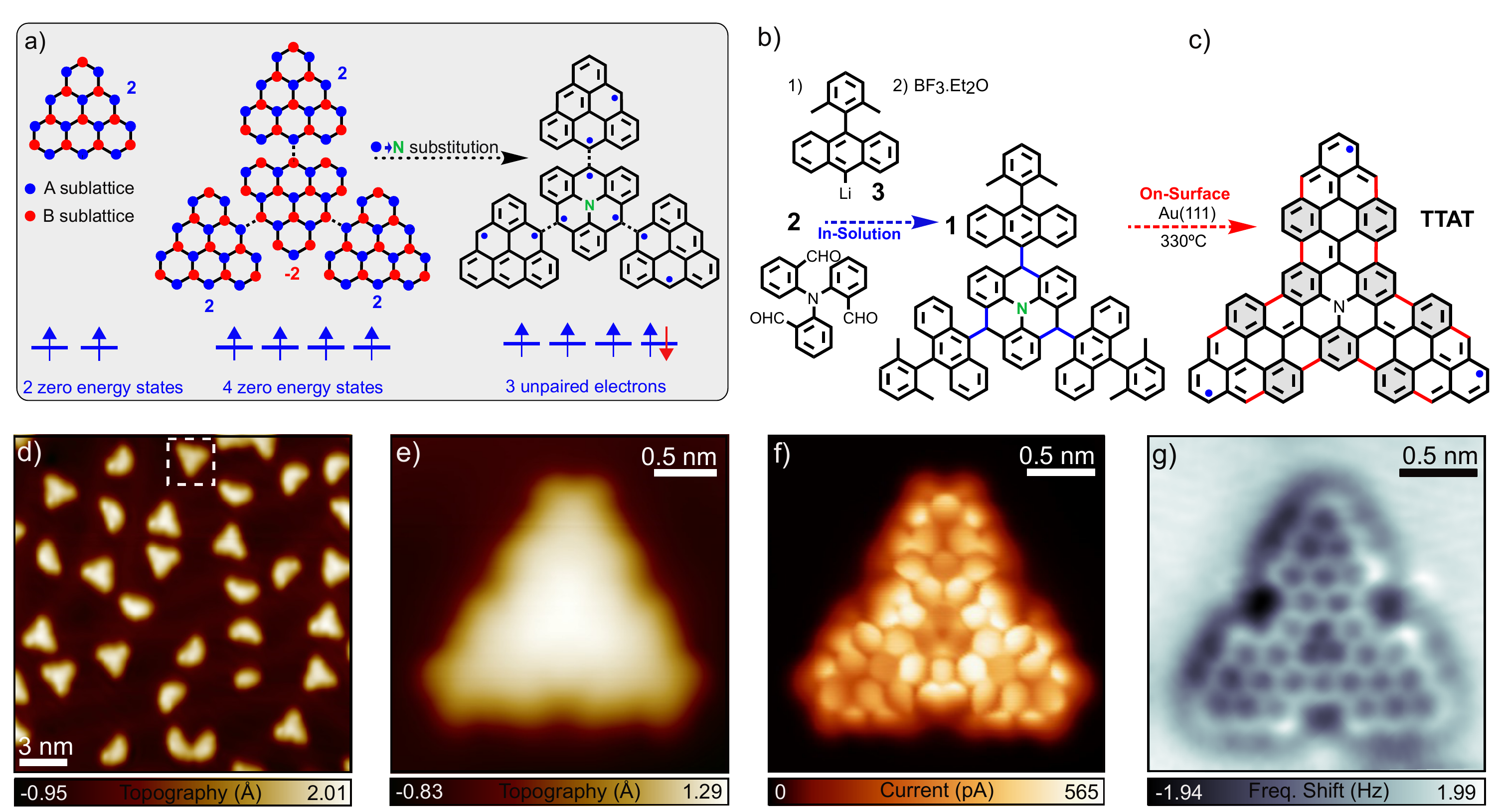}
\caption{(a) Schematic representation of the formation of the aza-triradical (\textbf{TTAT}) in (c) from four [3]triangulenes combined such that the majority sublattice of the central triangulene results opposite to the one of the three external triangulenes. Introducing an N atom in the center of the inner triangulene (in the majority sublattice) results in five electrons occupying four states. The final structure, therefore, hosts three unpaired electrons, expected to couple ferromagnetically by Hund's exchange. (b) In-solution and on-surface reaction steps leading to the synthesis of \textbf{TTAT}, with the C-C bonds formed during each step indicated in blue and red, respectively. (d) STM constant-current image ($V=0.9$ V, $I=30$ pA) after deposition of the precursor on Au(111) and subsequent annealing at 330 $^{\circ}$C. The white dotted square highlights an intact and planar molecule, corresponding to \textbf{TTAT}. (e) STM constant-current image of \textbf{TTAT} measured with a CO-functionalized tip ($V=200$ mV, I$=30$ pA). (f) Constant-height bond-resolved STM current scan ($V=5$ mV) and (g) constant-height bond-resolved AFM image (oscillation amplitude $A=60$ pm), performed with CO-functionalized tips.}
\label{fig:Fig1}
\end{figure*}

In this study, we implemented this connection strategy to synthesize a large, triangular nanographene with a symmetric trimer of ferromagnetically interacting localized radicals. By fusing three [3]triangulenes (3T) onto the edges of an aza[3]triangulene (A3T) core, we formed TTAT (tris-triangulene-aza-triangulene), a high-spin triradical nanographene that behaves as a Heisenberg spin triangle. Each 3T unit hosts two unpaired $\pi$ electrons localized along majority zigzag sites \cite{Pavlicek2017, Turco2023}. As shown in Fig.~\ref{fig:Fig1}a, this connection strategy pairs majority sublattice sites in opposite orientations, reducing the number of zero-energy states (i.e. their nullity \cite{Fajtlowicz2005, Wang2009a}) from eight to four and preserving four unpaired $\pi$ electrons for the pristine carbon structure. 

Within the A3T core, nitrogen substitution at a majority site introduces an additional electron to the $\pi$ system, stabilizing a $D_{3h}$ symmetric configuration with three unpaired electrons at the zigzag corners.
Structurally, TTAT resembles an aza[8]triangulene with six fewer six-membered rings, two per triangular side. This change creates three distinct gulf regions along the edges, each formed by two conjoined bay areas that accommodate nine Clar sextets  (see Fig.~\ref{fig:Fig1}c).  
 
In the following, we report the on-surface generation of \textbf{TTAT} on a Au(111) surface and demonstrate that this molecule behaves as a ferromagnetic Heisenberg spin triangle.  Combining low-temperature scanning tunneling microscopy (STM) measurements with theoretical simulations, we resolve its structural integrity on a surface and demonstrate that \textbf{TTAT} lies on a neutral charge state, maintaining a spin  3/2 ground state. The resolution of low-energy spectroscopic fingerprints and their simulation through multiconfigurational simulations revealed the presence of a triradical character with ferromagnetic interaction among its unpaired electrons. This spin triangle represents a unique system for investigating entanglement in a single-molecular architecture. 

\textbf{Synthesis strategy of TTAT:}  
We envisioned the synthesis of \textbf{TTAT} through a combination of in-solution and on-surface synthesis, as represented in Figs.~\ref{fig:Fig1}b and \ref{fig:Fig1}c, respectively. First, we addressed the preparation of the TTAT precursor \textbf{1} following the synthesis strategy of an aza-[5]-triangulene precursor \cite{Vilas-Varela2023} shown schematically in Fig.~\ref{fig:Fig1}b, and described in more detail in the Supplementary Section S\ref{sec:chemistry}.  Specifically, we followed a sequence of in-solution reaction steps based on 
the treatment of tribenzaldehyde \textbf{2} with an excess of organolithium \textbf{3}, followed by BF$_3$-promoted three-fold intramolecular Friedel-Crafts. Following this synthetic protocol, we isolated compound \textbf{1} in 38 \% yield. 

\begin{figure*}[t!]
\centering
\includegraphics[width=0.8\textwidth]{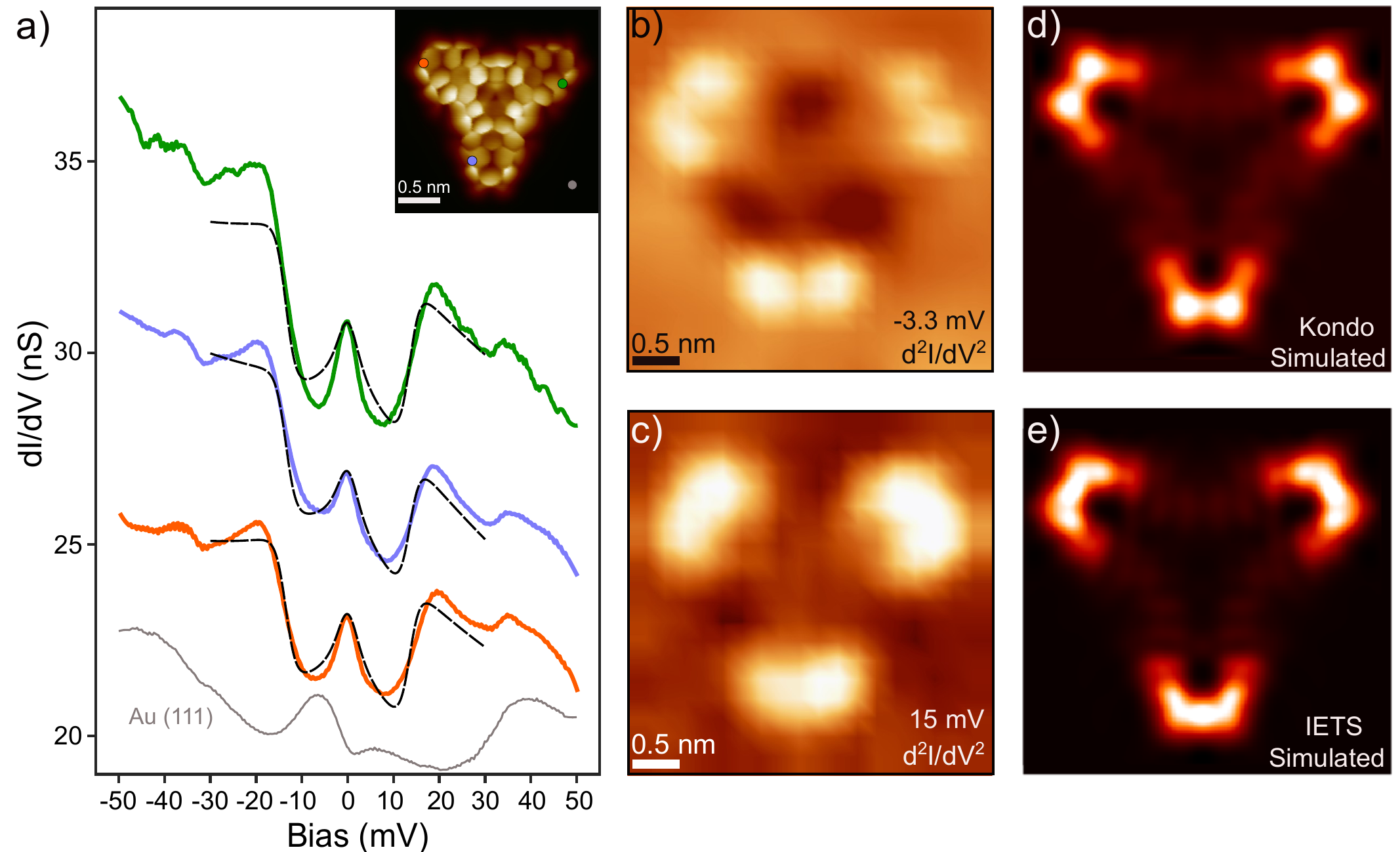}
\caption{(a) Low-energy $dI/dV$ spectra of \textbf{TTAT} measured with a CO-functionalized tip at the positions indicated in the inset. The spectra display a zero-bias resonance and inelastic spin excitation features at $V \approx$$ \pm 15$ mV. The black dashed lines represent fits to the data using the perturbative model by Ternes \cite{Ternes2015}, for the case of three $S=1/2$ spins coupled with a ferromagnetic exchange $J=9$ meV. Weaker steps at $\pm 35$ mV are attributed to the excitation of frustrated rotational modes of the CO molecule \cite{Delatorre2017}. Spectroscopy parameters: $V=50$ mV, $I=1$ nA, $V_\mathrm{mod}=2$ mV. (b,c) $d^{2}I/dV^{2}$ maps at $V=-3.3$ mV and $V=15$ mV, obtained by numerical differentiation from a grid of $dI/dV$ spectra. The maps probe the spatial distribution of the zero-bias resonance (b) and the inelastic signal (c) without elastic background effects. (d) Simulated Kondo and (e) spin excitation $dI/dV$ maps, computed from the Kondo orbitals (d) and Natural Transition Orbitals (NTOs), respectively (see text and Supplementary Information). }
\label{fig:Fig2}
\end{figure*}

We deposited the \textbf{TTAT} precursor \textbf{1} onto a Au(111) surface at room temperature via flash annealing of a silicon wafer loaded with molecular grains. 
Subsequently, the sample was annealed at 330$^{\circ}$C to activate the dehydrogenation reactions necessary to induce the formation of the 12 C-C bonds in red in Fig.~ \ref{fig:Fig1}c and culminate the molecular planarization. 
The overview STM constant-current image recorded after annealing (Fig.~\ref{fig:Fig1}d) shows intact triangular-shaped products alongside smaller molecular fragments. A closer inspection reveals that many triangular products retain one or more methyl groups, which appear as protruding rounded lobes in the STM images. Nevertheless, we identified the target product, \textbf{TTAT}, in a small fraction (around 5\%) of the non-fragmented molecules. The molecular structure appears fully planarized in this case and displays chamfered corners (Fig.~\ref{fig:Fig1}e). To conclusively demonstrate the successful on-surface generation of \textbf{TTAT}, we performed bond-resolved (BR) constant-height STM and non-contact AFM imaging using a CO-terminated tip \cite{Gross2009b} (Figs.~\ref{fig:Fig1}f-g). The BR images resolve the molecular backbone, revealing the absence of structural defects and the preserved three-fold symmetry of \textbf{TTAT} upon adsorption on the surface. Additionally, the bond-resolved STM image, recorded at $V=5$ mV, displays an apparent increase in the current signal along the zigzag edges near the triangulene corners and in the gulf regions around the center, providing a first indication of an enhanced density of states around the Fermi level, likely due to the presence of radical states \cite{Li2019}.

\begin{figure*}[t!]
   \includegraphics[width=\textwidth]{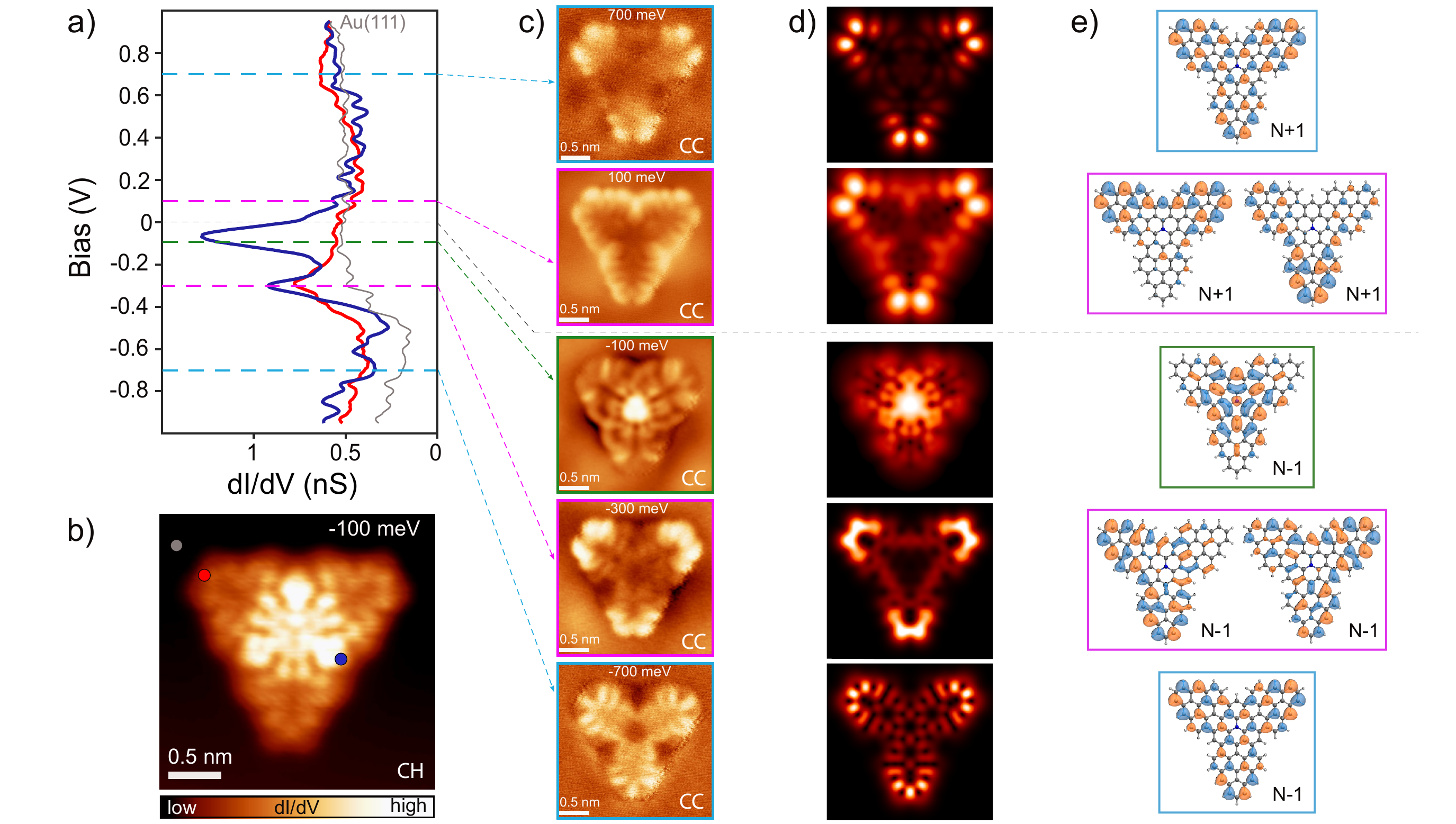}
	\caption{(a) $dI/dV$ spectra measured at the points indicated in b), revealing molecular orbital resonances ($V=1$ V, $I=500$ pA, $V_\mathrm{mod}=10$ mV). (b) Constant-height (CH) $dI/dV$ map recorded at $V=-100$ mV with a CO-functionalized tip, corresponding to an orbital with non-vanishing signal over the inner N-doped triangulene (open feedback parameters: $V=-100$ mV, $I=300$ pA, $V_\mathrm{mod}=10$ mV) . (c) Constant-current (CC) $dI/dV$ maps recorded at different bias values around 0, with a CO-terminated tip ($I=300$ pA, $V_\mathrm{mod}=10$ mV). (d) Simulated $dI/dV$ maps, obtained using Dyson orbitals, corresponding to the processes of adding and removing electrons. (e) Dyson orbitals isosurfaces. 
	}
\label{fig:Fig3}
\end{figure*}

\textbf{Evidence of open-shell states on Au(111):} 
To address the electronic and magnetic properties of \ttat\ on Au(111), we performed differential conductance ($dI/dV$) spectroscopy (Fig.~\ref{fig:Fig2}a). First, we explored the low-energy spectral window, where   spin fingerprints appear in the density of states. Spectra measured on the corners of the outer triangulenes show low-bias features indicative of an open-shell character: a weak zero-bias peak, characteristic of a Kondo resonance, and bias-symmetric step-like features at $V \approx$$ \pm 15$ mV, which are related to inelastic spin excitation processes \cite{Li2019,Ortiz2020,Mishra2020,Zheng2020,Hieulle2021,Mishra2021,Du2023,Krane2023e,Song2024,Turco2024}. 
To probe the spatial distribution of both Kondo and inelastic electron tunneling spectroscopy (IETS) features, we mapped the derivative of the differential conductance (i.e. $d^{2}I/dV^{2}$ maps) at $V=-3.27$ mV and $V=15$ mV, respectively. At these bias values, Kondo and IETS features appear as peaks in $d^{2}I/dV^{2}$ spectra, with amplitude proportional to the weight of the Kondo and inelastic channels \cite{Hieulle2021} (Fig.~\ref{figSI:FigS4}). As shown in Figs.~\ref{fig:Fig2}b and \ref{fig:Fig2}c, in both maps, the $d^{2}I/dV^{2}$ signal appears localized on the corners of the three outer triangulenes, indicating that the radical character of the molecule stems primarily from orbitals distributed over the edge of the external moieties, as we will discuss later.

Resolution of the \textbf{TTAT} frontier orbitals and their distribution over the molecular architecture provides a glimpse of the molecular spin ground state. We measured $dI/dV$ spectra on a broader bias range [1V,-1V] on distinct molecular positions (Fig.~\ref{fig:Fig3}a) and found several peaked resonances attributed to molecular states. The most protruding one is a clear resonance centered at around $-100$ mV, with spatial distribution over the central aza moiety, as probed by the constant-height $dI/dV$ map reported in Fig.~\ref{fig:Fig3}b. 
The tail of this resonance crosses through zero bias and causes the tilted background in the low-energy spectra of Fig.~\ref{fig:Fig2}a. It also accounts for the increased current around the center observed in bond-resolved constant-height images like in Fig.~\ref{fig:Fig1}f. 
As shown in Fig.~\ref{fig:Fig3}c, a non-vanishing signal over the central aza moiety is only found in the $dI/dV$ map measured at $-100$ meV, with no replica at positive bias. This points towards a doubly-occupied state over the center of the flake, in agreement with results from DFT calculations reported in Supplementary Fig.~\ref{figSI:DFT} for a free molecule. According to DFT, the zero-energy orbital A$_1$ (with the largest amplitude over the N site and C$_{3v}$ symmetry) hosts the extra electron provided by the N heteroatom substitution in the neutral charge state. 
The molecule spin is primarily hosted in the three remaining singly-occupied orbitals (SOMOs), two of which are degenerated, with E symmetry, and the third one is the C$_{3v}$ symmetric A$_2$ orbital \cite{Sandoval-Salinas2019}. These orbitals are distributed mainly along the molecular edges, with a weak contribution from the aza group. 

\begin{figure*}[t!]
\centering
\includegraphics[width=\textwidth]{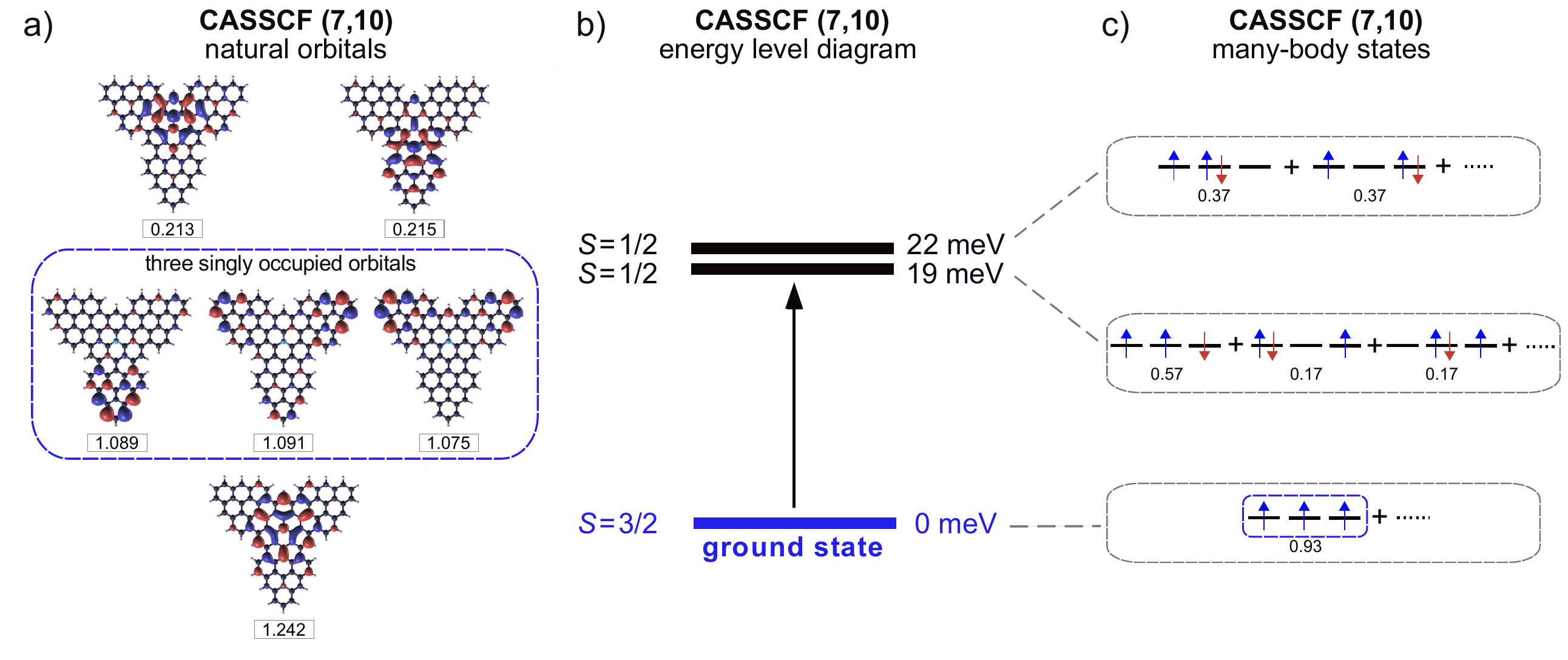}
\caption{(a) Natural orbitals computed by CASSCF(7,10). The numbers at the bottom of each orbital indicate the electron occupation. We find three orbitals with occupation close to 1, spatially distributed over the triangulene edges. (b) Diagram representing the energy and the total spin of the many-body ground state and first two excited states of \textbf{TTAT}, as computed by CASSCF(7,10). (c) Schematic representation of the most relevant Slater determinants for each of the many-body states in (b), displaying the electronic occupation of the three natural orbitals highlighted in the dashed box in (a). The number below each Slater determinant refers to its weight in the corresponding many-body state.}
\label{fig:Fig4}
\end{figure*}
 
Their singly occupied character can be concluded from  $dI/dV$ maps throughout a wider energy region: characteristic $dI/dV$ patterns attributed to the SOMOs and their correlated singly unoccupied orbitals (SUMOs) appeared at $-700$ meV and $-300$ meV, and at 100 meV and 700 meV, respectively.  Since all states detected around \ef\ lie close in energy, the molecular system is expected to exhibit a strong multiconfigurational character. Therefore, to identify and interpret the $dI/dV$ maps, we computed the relevant Dyson orbitals \cite{Ortiz2020a} using the natural orbitals obtained from Complete Active Space Configuration Interaction (CASCI) calculations (see Methods and Supplementary Information). In agreement with DFT, we obtained three Dyson orbitals accounting for electron addition and four for electron removal (Fig.~\ref{fig:Fig3}e), as expected for a ground state composed of three singly occupied and one doubly occupied state. In Fig.~\ref{fig:Fig3}d, we show the simulated $dI/dV$ maps resulting from the computed Dyson orbitals, including a CO-functionalized tip, calculated with the PP-STM code \cite{Krejci2017}. The maps reproduce the experimental $dI/dV$ maps in great detail, further confirming the identification of three singly occupied states hosting the spin properties of the molecule.

The simulations indicate that \textbf{TTAT} maintains a neutral charge state on the electrophilic Au(111) surface. 
Kelvin probe force microscopy (KPFM) measurements provided in Supplementary Fig.~\ref{figSI:FigS5} confirm that the molecule remains in a neutral state on the Au(111) substrate. This behavior contrasts with the cationic state found for the structurally similar molecule aza-[5]-triangulene (A5T), despite both molecules having the same spin imbalance and nullity \cite{Vilas-Varela2023,Lawrence2023,CalvoFernandez2024}. 
Although the precise factors driving charge transfer differences near chemical equilibrium would require further study, we speculate that the neutral stability of \textbf{TTAT} (compared to \textbf{TTAT$^+$}) is related to its enhanced aromaticity,  evidenced, for example, by the larger number of Clar sextets.
In contrast, the antiaromaticity of neutral A5T species accounts for its tendency to oxidize in Au(111) \cite{Vilas-Varela2023}.  

These observations suggest that \textbf{TTAT} has a spin ground state $S=3/2$ on Au(111), with parallel spin alignment due to Hund’s exchange interactions among the three SOMOs \cite{Jacob2022a,Mishra2022a}.
We thus attribute the zero-bias resonance in the STS spectra to an underscreened Kondo effect associated with this high-spin state.
Nanographenes with spin above $S=1/2$ generally exhibit partial Kondo screening on surfaces, leading to smaller zero-bias resonances \cite{Li2020,Turco2023}, as seen in Fig.~\ref{fig:Fig2}a. Additional evidence for this underscreened Kondo effect is the resonance splitting under an external magnetic field ($B=2.7$ T) shown in Supplementary Fig.~\ref{figSI:FigS6}, supporting an $S=3/2$ ground state for TTAT on Au(111).

\textbf{Triradical character of TTAT:}
To rationalize the high-spin triradical ground state  of \textbf{TTAT}, and the origin of the IETS features observed in $dI/dV$ spectra, we performed multiconfigurational calculations with the complete active space self-consistent (CASSCF) method. Considering a CAS(7,10), we obtain that three natural orbitals (highlighted in the dashed box, Fig.~\ref{fig:Fig4}a) have an electron occupation close to 1 in the ground state, thus indicating the presence of three unpaired electrons. In accordance with the DFT calculations and the experimental results, a natural orbital with A$_1$ symmetry and centered on the N heteroatom appears with a larger electron occupation. 
As shown in the many-body energy levels in Fig.~\ref{fig:Fig4}b, the CASSCF calculations confirm the quartet ($S=3/2$) ground state of \textbf{TTAT}, which accordingly is predominantly described by a single determinant \cite{Krylov2005}, as depicted in Fig.~\ref{fig:Fig4}c.

The first two excited states are two nearly degenerate spin $S=1/2$ levels found at 19 meV and 22 meV above the ground state. These values lie close to the experimental excitation gap obtained from the IETS steps ($\Delta \sim$ $15$ meV), supporting the identification of the inelastic spectral features with a spin excitation process from $S=3/2$ to $S=1/2$. These doublets are described by a linear combination of different Slater determinants with similar weight (Fig.~\ref{fig:Fig4}c), justifying multiconfigurational methods utilized in Fig.~\ref{fig:Fig3}. 

\begin{figure*}[htb]
\includegraphics[width=0.7\textwidth]{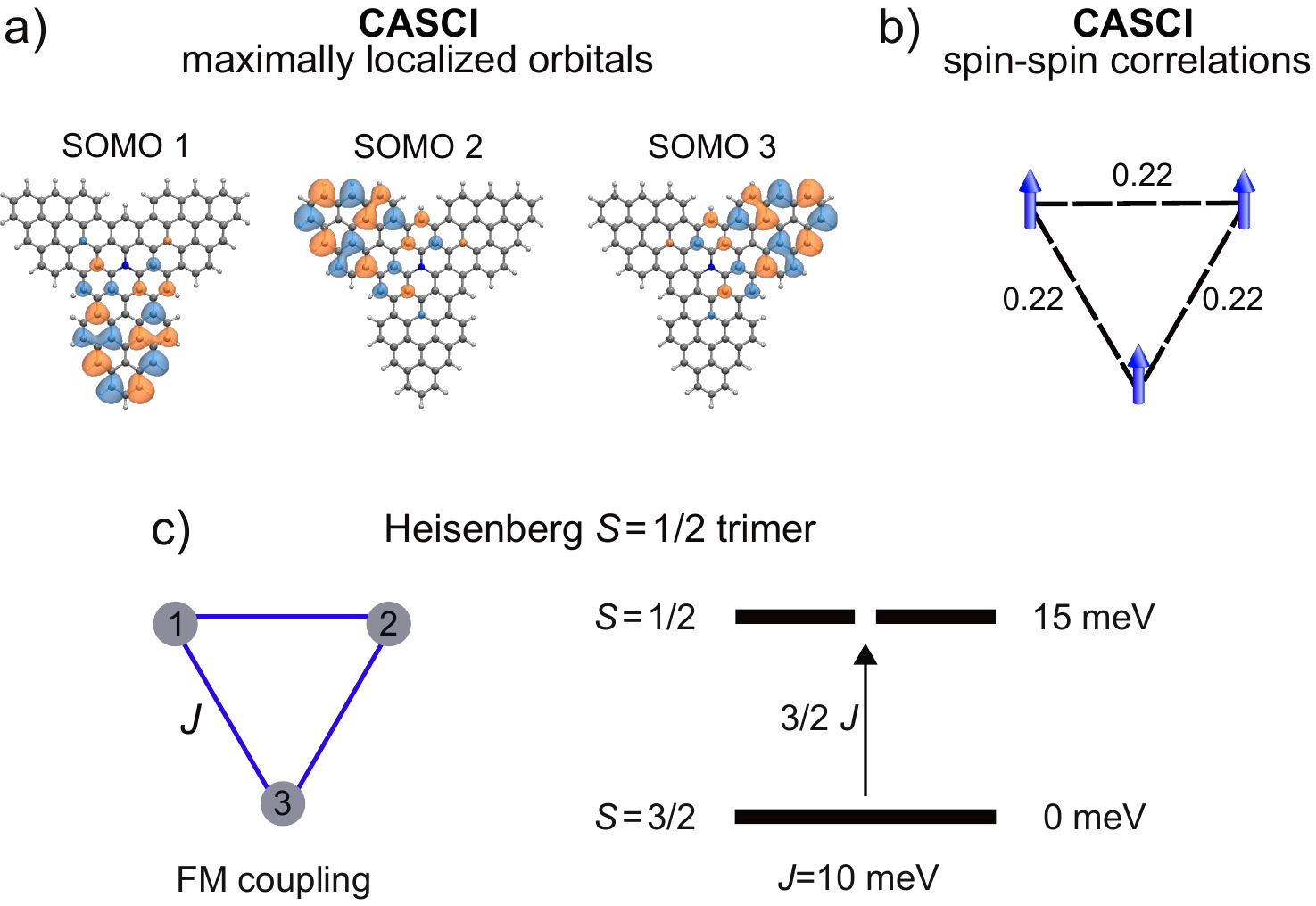}
\caption{(a) Representation of the three singly-occupied orbitals using a maximally localized basis set, which shows that each spin is mostly located on a triangulene corner. (b) Spin-spin correlation between each pair of spins computed using the orbital representation in (a).
This picture describes the magnetic state of \textbf{TTAT} in terms of a symmetric Heisenberg ferromagnetic trimer, as illustrated in (c). According to this model the experimental quartet-doublet energy gap of $15$ meV corresponds to an exchange coupling $J=10$ meV \cite{Haraldsen2005}.
	}
\label{fig:Fig5}
\end{figure*}

The multiconfigurational nature of the doublet excited states is likely reflected in the experimental magnetic fingerprints shown in  Figs.~\ref{fig:Fig2}b-\ref{fig:Fig2}c. To explain the spatial localization of the Kondo signal, we used the concept of Kondo orbitals recently introduced in Ref.~\onlinecite{CalvoFernandez2024} to describe the Kondo effect in open-shell polyradical molecular systems. In this framework, the Kondo orbitals are associated with scattering processes between molecular electrons and conduction electrons of the underlying metal featuring antiferromagnetic exchange coupling. Using CASCI calculations, we obtained the set of Kondo orbitals shown in Supplementary Fig.~\ref{figSI:FigS9}, which we used for calculating the corresponding Kondo $dI/dV$ map using the PP-STM code \cite{Krejci2017}. The simulated  $dI/dV$ map shown in Fig.~\ref{fig:Fig2}d reproduces the shape and distribution of the experimental Kondo map, further confirming the interpretation of the Kondo signal as the fingerprint of a $S=3/2$ spin state.

Similarly, the inelastic spin excitation from the quadruplet ground state to the two doublet excited states is represented by CASCI Natural Transition Orbitals (NTOs) shown in  Supplementary Fig.~\ref{figSI:FigS8}. The simulated spin excitation $dI/dV$ map \cite{Krejci2017} reported in Fig.~\ref{fig:Fig2}e was obtained by summing the contributions of the NTOs corresponding to the spin excitation to the two degenerate excited states. The excellent agreement with the experimental map of the IETS signal supports the origin of the inelastic signal as a quartet-doublet spin excitation.

The localization of spin fingerprints at the three vertices suggests that the open-shell character of \textbf{TTAT} could be described by three spatially localized radicals rather than a set of overlapping SOMOs. Using a maximally localized orbital basis set, we obtain a representation where each of the three unpaired electrons is mostly located at an individual triangulene corner, as displayed in the SOMOs in Fig.~\ref{fig:Fig5}a. We computed the spin-spin correlation $A_{ij} = \langle\hat{S_{i}}\hat{S_{j}}\rangle -\langle\hat{S_{i}}\rangle \langle\hat{S_{j}}\rangle$ for each pair of spins $i$ and $j$ in those maximally localized orbitals. The results, illustrated in Fig.~\ref{fig:Fig5}b, confirm the ferromagnetic coupling between three unpaired spins in the ground state with the value of the spin-spin correlation $A_{ij} =0.22$ au. 

This representation suggests the possibility of describing the triradical molecule \textbf{TTAT} as a symmetric Heisenberg $S=1/2$ trimer \cite{Haraldsen2005}. Considering an equal ferromagnetic coupling $J$ between the three unpaired spins (Fig.~\ref{fig:Fig5}b), a Heisenberg spin model yields a ground state with total spin $S=3/2$ and two degenerate doublets ($S=1/2$) as first excited states, similarly to the results of the many-body CASSCF calculations. In the case of an equilateral Heisenberg trimer, the exchange $J$ is given by $J=\frac{2}{3}\Delta E$, where $\Delta E$ is the quartet-doublet energy difference. Therefore, considering the experimental excitation energy $\Delta E = 15$ meV, we determine for \textbf{TTAT} an exchange coupling $J=10$ meV, in good agreement with the value of 9 meV obtained from the fit to the STS data using the perturbative model by Ternes \cite{Ternes2015}. 


\section{Conclusions}
In summary, we have presented a polyradical aza-nanographene (\textbf{TTAT}) hosting three unpaired $\pi$ electrons localized at the vertices of a triangle and coupled through symmetric, ferromagnetic interactions. It was designed by combining well-known molecular building blocks, all-carbon and N-doped [3]triangulenes, and fabricated via a combination of in-solution and on-surface synthesis. The detection of clear magnetic fingerprints in scanning tunneling spectroscopy (i.e., both a weak Kondo resonance and a IETS step-like feature) demonstrated the open-shell and polyradical character of the molecule on Au(111). Combining differential conductance spectra and orbital maps with DFT and advanced multi-configurational CASSCF calculations, we revealed the presence of three radicals and their ferromagnetic alignment, resulting in a $S=3/2$ (quartet) ground state. The evidence of multi-radical interactions and the high-spin configuration make this system a potential candidate for applications in carbon-based spintronics and a promising platform for the exploration of magnetic states in extended molecular structures.

\section*{Methods}
The (111) surface of a gold single crystal was cleaned by several cycles of sputtering with Ne$^+$ ions and subsequent annealing at $T=600$°C under ultrahigh vacuum (UHV) conditions. The precursor of \textbf{TTAT} was prepared in solution as described in Fig.~\ref{fig:Fig1} and in the Supplementary Information (Section \ref{sec:chemistry}). The sublimation was achieved via fast thermal heating of a Si wafer loaded with grains of compound \textbf{1}. 

All measurements were conducted in a low-temperature STM at $5$~K in ultrahigh vacuum (UHV) conditions, except for those reported in Fig.~\ref{figSI:FigS6}, performed in a commercial Joule-Thompson (JT) STM with a base temperature of 1.2~K.
STM constant-current images were performed with a gold-coated tungsten tip or, when indicated in the text, with a CO-terminated tip. The bond-resolved STM and AFM images were always recorded using a CO-functionalized tip in constant-height mode.
The figures representing the experimental data were prepared using the WSxM and SpectraFox software \cite{Horcas2007, Ruby2016}. 

Differential conductance spectra were recorded using a lock-in amplifier with frequency $f=753$ Hz ($f=887$ Hz for the spectra in Fig.~\ref{figSI:FigS6}). The modulation amplitude and current parameters are indicated in the captions of the respective figures. The $d^{2}I/dV^{2}$ maps reported in Fig.~\ref{fig:Fig2} were obtained by numerical differentiation from a grid of $dI/dV$ spectra made up of 20x20 points, using the WSxM software \cite{Horcas2007}. 

The AFM bond-resolved image reported in Fig.~\ref{fig:Fig1} and the Kelvin probe force microscopy measurements reported in Fig.~\ref{figSI:FigS5} were performed using a qPlus-type sensor with eigenfrequency $f_{0}=30.72$ kHz and $Q$-factor of the order of $10^{4}$ \cite{Giessibl1998}. The AFM was operated in the frequency modulation mode \cite{Albrecht1991}, with an oscillation amplitude $A=60$ pm. 

 
\section*{Acknowledgments}
The authors acknowledge financial support from grants PID2022-140845OBC61, PID2022-140845OBC62, PID2022-139776NB-C65, PID2023-146694NB-I00, and CEX2020-001038-M funded by MICIU/AEI/ 10.13039/501100011033 and the European Regional Development Fund (ERDF, A way of making Europe), from the FET-Open project SPRING (863098), the HE project HORIZON-EUROHPC-JU-2021-COE-01-01-101093374-MaX,  the ERC Synergy Grant MolDAM (no.~951519), and the ERC-AdG CONSPIRA (101097693) funded by the European Union, from projects 2022-QUAN-000030-01 and 2023-QUAN-000028-01 funded by the Diputacion Foral de Gipuzkoa, 
from the European Union NextGenerationEU / PRTR-C17.I1 as well as by the IKUR Strategy of the Department of Education of the Basque Government, and from the Xunta de Galicia (Centro singular de investigación de Galicia accreditation 2019-2022, ED431G 2019/03 and Oportunius Program).
A.V. and F.R.-L. acknowledge enrollment in the doctorate program “Physics of Nanostructures and Advanced Materials” from the advanced polymers and materials, physics, chemistry and technology” department of the Universidad del País Vasco (UPV/EHU).  F.R.-L. acknowledges funding by the Spanish Ministerio de Educación y Formación Profesional through the PhD scholarship No. FPU20/03305. F.S. acknowledges funding by the Spanish Ministerio de Ciencia, Innovación y Universidades through the Ramón y Cajal Fellowship RYC2021-034304-I. P.J., M.K. and D.S. acknowledge the support of the Czech Science Foundation (GACR) project No.23-05486S and the CzechNanoLab Research Infrastructure supported by MEYS CR (LM2018110). E.A.~acknowledges EPSRC grant EP/V062654/1 from the Theoretical Condensed Matter Cambridge - Critical Mass Grant United Kingdom and computational resources provided by the Donostia International Physics Center (DIPC) Computer Center.

\providecommand{\latin}[1]{#1}
\makeatletter
\providecommand{\doi}
  {\begingroup\let\do\@makeother\dospecials
  \catcode`\{=1 \catcode`\}=2 \doi@aux}
\providecommand{\doi@aux}[1]{\endgroup\texttt{#1}}
\makeatother
\providecommand*\mcitethebibliography{\thebibliography}
\csname @ifundefined\endcsname{endmcitethebibliography}  {\let\endmcitethebibliography\endthebibliography}{}

\end{document}


\date{\today}
\renewcommand{\abstractname}{\vspace{1cm} }

\twocolumn[
\begin{@twocolumnfalse}
\oldmaketitle
\begin{abstract}
\setstretch{0.9}
\vspace{2mm}
	\tableofcontents		
\end{abstract}	
\end{@twocolumnfalse}
]
 	

\onecolumn
\section{Synthetic details}
\label{sec:chemistry}
\subsection{Solution Synthesis of Molecular Precursors}

Starting materials were purchased reagent grade from TCI and Sigma-Aldrich and used without further purification. Trimethyl 2,2',2''-nitrilotribenzaldehyde (\textbf{2}) was obtained following reported procedures \cite{Vilas-Varela2023}. Organolithium reagent \textbf{3} was prepared by treatment of the corresponding bromide\cite{Mishra2021b} with n-butyllithium. All reactions were carried out in flame-dried glassware under an inert atmosphere of purified Ar using Schlenk techniques. Thin-layer chromatography (TLC) was performed on Silica Gel 60 F-254 plates (Merck). Column chromatography was performed on silica gel (40-60 µm). NMR spectra were recorded on a Bruker Varian Inova 500 spectrometer.

\subsection{Synthesis of the TTAT Precursor}

\begin{figure*} [hbt]
\centering
    \includegraphics[width=0.6\columnwidth]{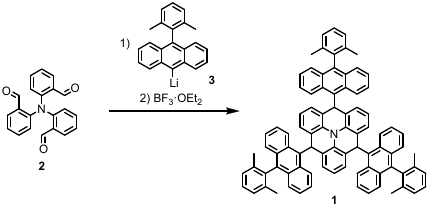}
\caption{Synthesis of the \textbf{TTAT} precursor, compound \textbf{1}.
	}
 \label{figSI:NMR1} 
\end{figure*}

Over a solution of organolithium derivative \textbf{3} (1.74 mmol, 5.00 equiv) in THF (30 mL), aldehyde \textbf{2} (115 mg, 0.35 mmol) was added at 0 ºC. After addition, the mixture was allowed to reach room temperature and stirred for 16 h. MeOH (2 mL) was then added, and the solvents were evaporated under reduced pressure. The residue was redissolved in CH$_2$Cl$_2$ (15 mL) and BF$_3\cdot$OEt$_2$ (1.00 mL) was added at 0 ºC. After addition, the mixture was stirred for 20 min and then poured into an aqueous NaOH solution (0.50 M, 40.0 mL). After separation of phases, the aqueous fraction was extracted with CH$_2$Cl$_2$ (2 x 15 mL). The combined organic fractions were dried over anhydrous Na$_2$SO$_4$, filtered, and evaporated under reduced pressure. The residue was purified by column chromatography (SiO$_2$; hexane:CH$_2$Cl$_2$, 4:1), affording compound 1 (85 mg, 38\%) as a yellow solid.

\subsection{NMR Data}

\textbf{$^1$H NMR (500 MHz, CDCl$_3$)} $\delta$: 8.69 – 8.59 (m, 3H), 7.55 – 7.48 (m, 3H), 7.47 – 7.36 (m, 5H), 7.32 – 7.26 (m, 3H), 7.21 – 7.14 (m, 5H), 6.52 – 6.47 (m, 2H), 6.43 (d, J = 7.7 Hz, 1H), 6.36 (t, J = 7.7 Hz, 1H), 1.73 (s, 3H), 1.71 (s, 3H), 1.63 (s, 3H) ppm. 

\textbf{$^{13}$C NMR (125 MHz, CDCl$_3$)} $\delta$: 138.72 (C), 138.48 (C), 138.20 (C), 138.13 (C), 137.91 (C), 137.87 (C), 136.76 (C), 136.71 (C), 133.13 (C), 132.92 (C), 132.76 (C), 132.22 (C), 132.01 (C), 131.99 (C), 131.10 (C), 131.04 (C), 129.23 (C), 129.12 (C), 129.10 (C), 128.25 (CH), 128.13 (CH), 127.83 (CH), 127.64 (CH), 127.61 (CH), 127.25 (CH), 127.23 (CH), 127.21 (CH), 127.17 (CH), 127.12 (CH), 127.03 (CH), 126.05 (C), 125.65 (C), 125.62 (CH), 125.58 (C), 125.43 (CH), 125.40 (CH), 125.34 (CH), 125.22 (CH), 125.19 (CH), 123.89 (CH), 123.72 (CH), 123.38 (CH), 123.29 (CH), 39.16 (CH), 20.30 (CH$_3$), 20.27 (CH$_3$), 20.18 (CH$_3$) ppm. 
\textbf{MS (APCI)} m/z (\%): 1122 (M+1, 93), 1019 (48), 841 (58). \textbf{HRMS (APCI)}: C$_{87}$H$_{64}$N; calculated: 1122.5033, found: 1122.5011.

\begin{figure*} [hbt]
    \includegraphics[width=0.8\columnwidth]{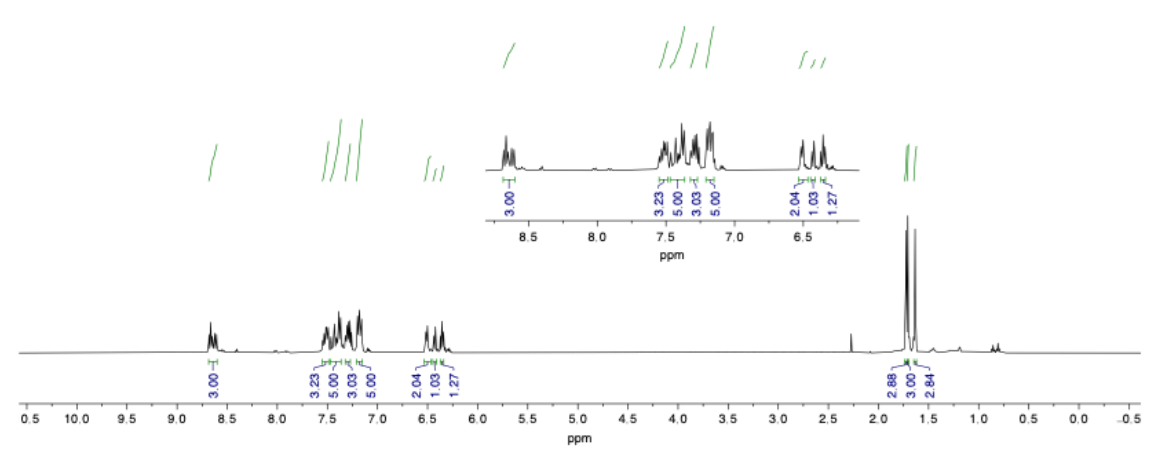}
\caption{$^1$H NMR (500 MHz, CDCl$_3$) spectrum of compound \textbf{1}.
	}
 \label{figSI:NMR1} 
\end{figure*}

\begin{figure*} [hbt]
    \includegraphics[width=0.8\columnwidth]{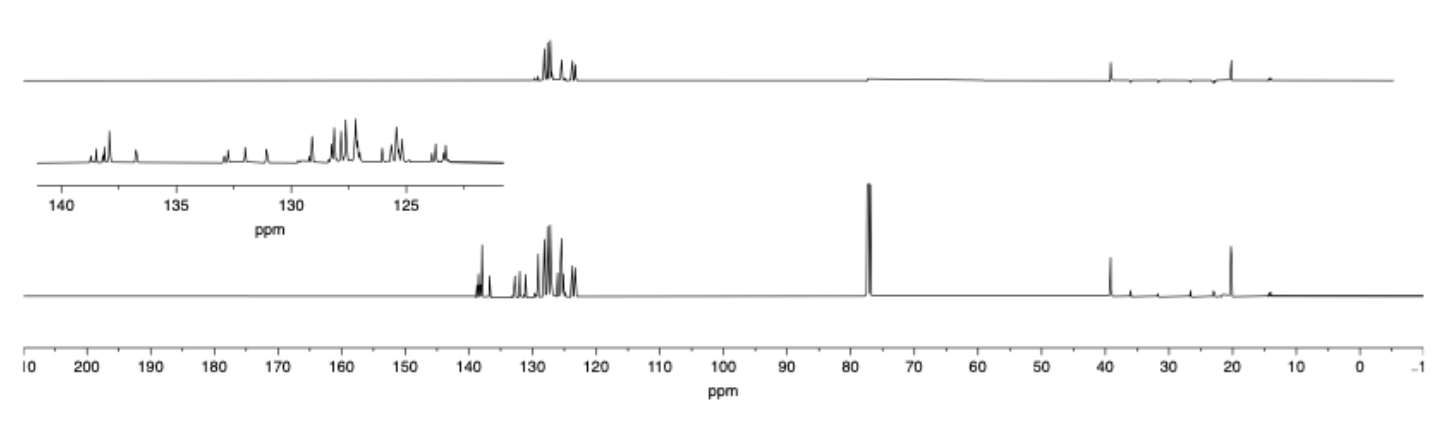}
\caption{$^{13}$C NMR-DEPT (125 MHz, CDCl$_3$) spectra of compound \textbf{1}.
	}
 \label{figSI:NMR2} 
\end{figure*}

\newpage

\section{Complementary experimental data}


\begin{figure*} [hbt]
    \includegraphics[width=\columnwidth]{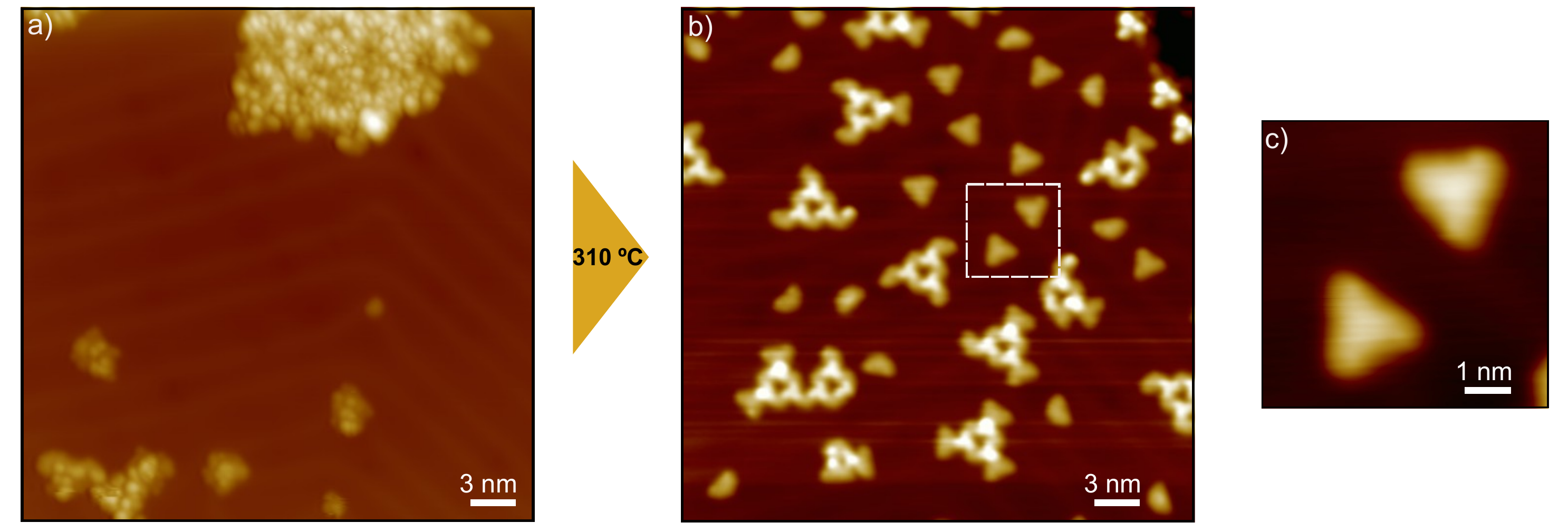}
\caption{a) STM constant-current overview image ($V=1$ V; $I=30$ pA) after deposition of the molecular precursor on Au(111) at room temperature, showing mostly three-dimensional molecular clusters and  aggregates. b-c) After annealing at 310$^{\circ}$C, smaller domains and individual molecules are observed. The presence of rounded, protruding corners indicate that the isolated molecules have not undergone the cyclodehydrogenation reaction yet, and thus still retain their methyl groups ($V=1$ V; $I=30$ pA). A further annealing at 330$^{\circ}$C is needed in order to fully activate the on-surface reaction and generate \textbf{TTAT} (Fig.~1 in the main text).
	}
 \label{figSI:FigS2} 
\end{figure*}

\begin{figure*} [hbt]
    \includegraphics[width=\columnwidth]{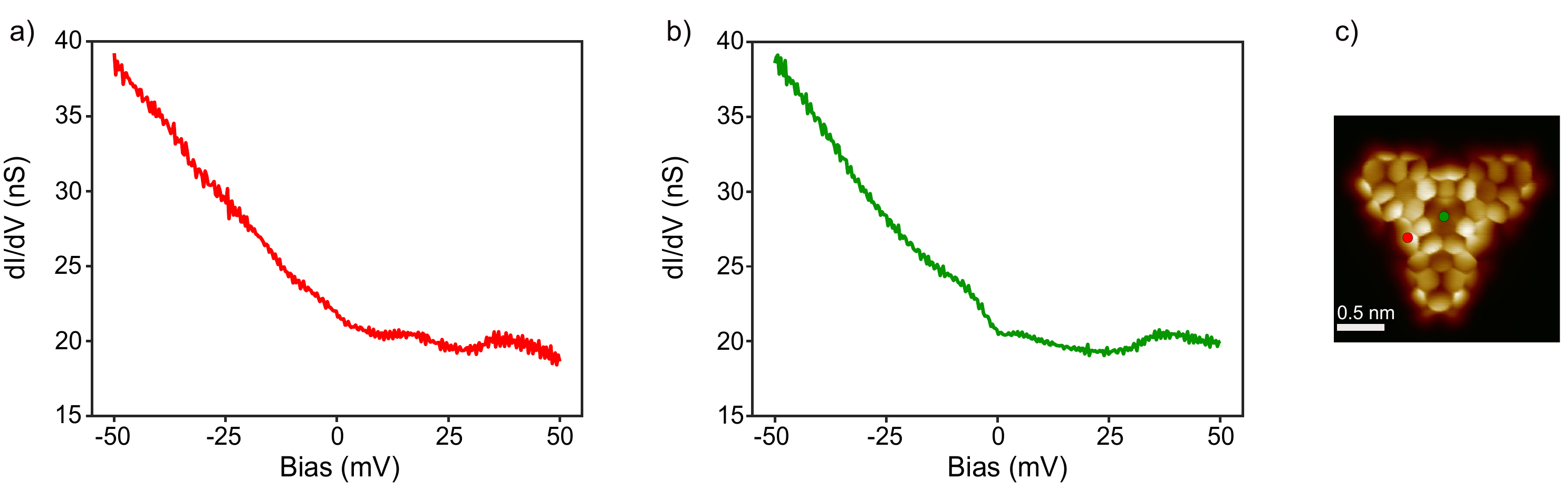}
\caption{Low-bias $dI/dV$ spectra measured with a CO-functionlized tip on the gulf area between the external triangulenes (a) and on the central N atom (b), as indicated in the bond-resolved constant-height image ($V=5$ mV) in c). The Kondo and spin excitation features measured on the triangulene corners (Fig.~2 in the main text) are absent in these regions, while we observe here a conductance increase due the onset of the molecular orbital at $V=-100$ mV. Spectroscopy parameters: $V=50$ mV, $I=1$ nA, $V_{mod}=2$ mV.}
 \label{figSI:FigS3} 
\end{figure*}

\begin{figure*} [hbt]
    \includegraphics[width=\columnwidth]{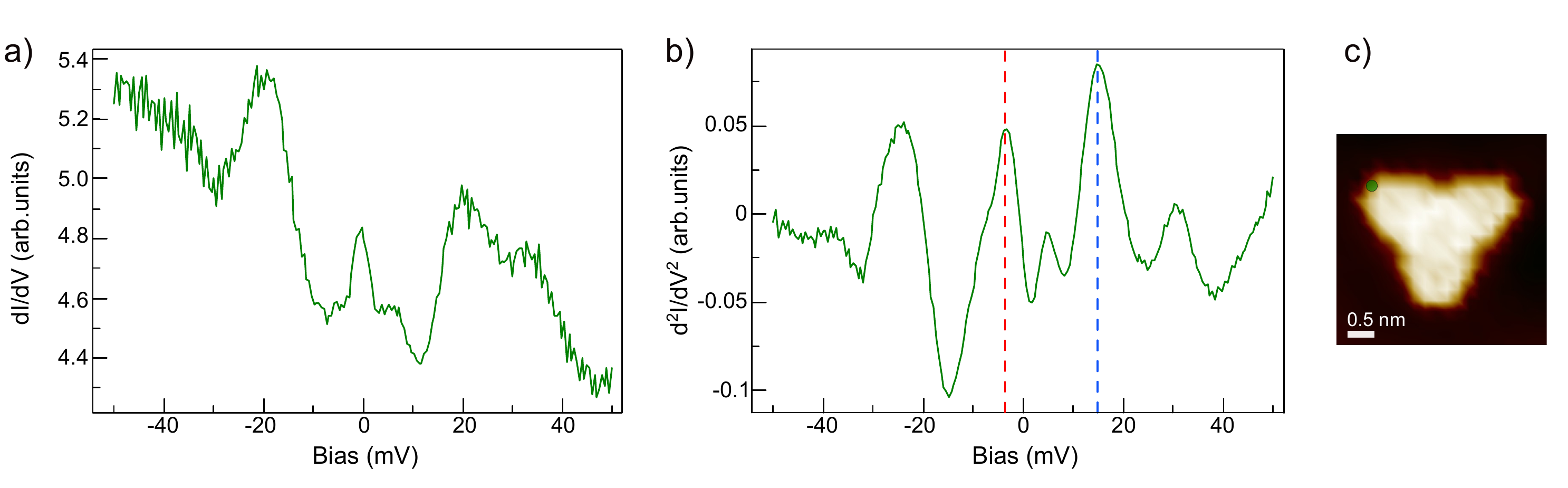}
\caption{Single $dI/dV$ spectrum (a) and correspondent $d^{2}I/dV^{2}$ plot (b) extracted from a $dI/dV$ grid in the position indicated in c). Alongside the Kondo and spin excitation features, we can observe inelastic steps around $V=\pm 5$ mV and $V=\pm 35$ mV that originate from vibrational modes of the CO molecule attached to the tip. The red and blue dashed lines in (c) indicate the peaks (at $V=-3.3$ mV and $V=15$ mV) that were selected to map the spatial distribution of the Kondo and the spin excitation feature, respectively, as reported in Fig.~2 in the main text. Parameters: $V=50$ mV, $I=1$ nA, $V_{mod}=2$ mV.}
 \label{figSI:FigS4} 
\end{figure*}

\begin{figure*} [hbt]
    \includegraphics[width=\columnwidth]{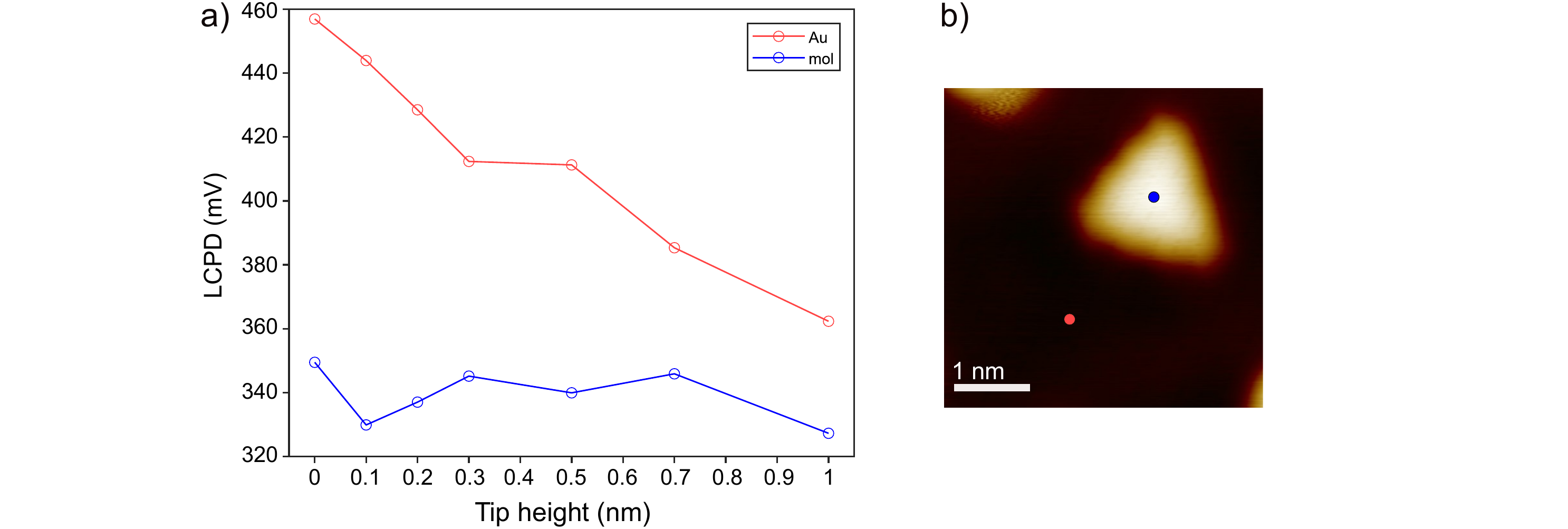}
\caption{Kelvin probe force microscopy (KPFM) measurements on \textbf{TTAT}. a) Local contact potential difference (LCPD) values measured at different tip-molecule distances on the molecule and on the bare Au(111), as indicated in the constant-current STM image in (b). The 0 distance corresponds to the setpoint ($I=30$~pA; $V=200$ mV) over the centre of the molecule. For each tip height, we first measure a $\Delta f(V)$ spectrum (frequency shift as a function of the tip-sample bias) in the range (-0.1V, 0.9V) over the molecule, then we move the tip to the bare Au (with the feedback loop open) and record the same spectrum at the same height, and determine the LCPD for each position from the vertex of the KPFM parabola \cite{Gross2009}.
We observe that the LCPD values on the molecule are always more negative than those measured on the substrate. However, this lower LCPD value is not due to a net charge of the molecule, but to the so-called \textit{push-back} effect (also known as pillow effect). This effect arises when a molecule (or an atom) is adsorbed on a metallic surface, and refers to the compression of the electron density of the metal that leaks into the vacuum by the adsorbate. It reduces the local work function of the surface \cite{Ishii1999} and gives rise to a small shift of the LCPD towards more negative bias voltages \cite{Trishin2022}. This is a general characteristic of metal-organic interfaces and independent of any charge transfer between adsorbate and metal surface. If, in addition, there are changes in the adsorbate´s charge state, i.e., net charges on the molecule, this would further modify the local work function. In that case, the shift of the LCPD would be much more pronounced (typically a few hundred meV at close tip-sample distances) and would show a strong dependence on the tip-sample distance \cite{Gross2009, Leoni2011, Gao2023}. Because these effects are not observed in our LCPD measurements, we conclude that charge transfer between TTAT and the Au(111) surface is negligible.}
 \label{figSI:FigS5} 
\end{figure*}

\begin{figure*} [hbt]
    \includegraphics[width=\columnwidth]{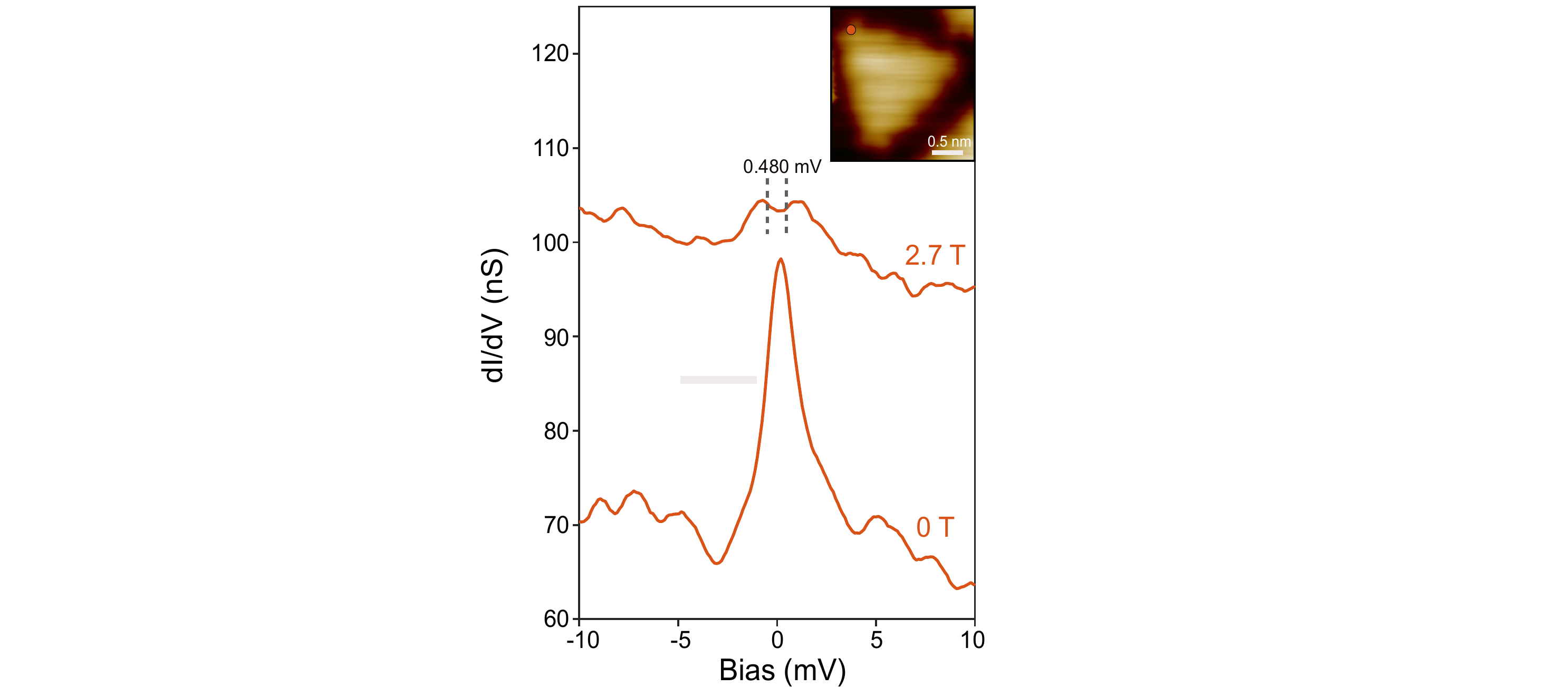}
\caption{Comparison of low-bias $dI/dV$ spectra measured in the position indicated in the inset, in the absence of external magnetic field, and at 2.7 T field. The Kondo resonance shows a splitting in the presence of the magnetic field, which can be interpreted as an indication of an underscreend Kondo effect. 
An underscreened Kondo resonance is expected to undergo a splitting as soon as the Zeeman energy ($g\mu_{B}B$) becomes greater than the thermal broadening ($k_{B}T$), which in our experiment ($T=1.2$ K) corresponds to a field $B \approx 1$T \cite{Li2019}. A fully screened $S=1/2$ Kondo resonance, on the other hand, would not display any splitting at this relatively low field, as we showed in previous works \cite{Vilas-Varela2023,Li2020}.
This observation suggests a high-spin ground state (i.e., larger than 1/2), but does not allow determining the exact total spin number \cite{Vilas-Varela2023}. However, considering the neutral charge state of \textbf{TTAT} on Au(111), deduced from the dI/dV maps of Fig.~3 and the KPFM data of Fig.~\ref{figSI:FigS5}, the measurements are consistent with a S=3/2 ground state \cite{Vilas-Varela2023,Li2020}.  Parameters: $V=10$ mV, $I=1$ nA, $V_{mod}=0.5$ mV. These measurements were performed at $T=1.2$ K. The markers used to determine the energy splitting of the resonance are placed at the points of highest slope within the split peak, following the procedure used in our previous works \cite{Vilas-Varela2023,Li2020}. It is important to note, however, that estimating the exact value of the energy splitting is not straightforward for weak magnetic fields and in the presence of significative overshoots (third order effects) \cite{Zhang2013}.}

\label{figSI:FigS6} 
\end{figure*}

\clearpage

\section{Complementary theoretical methods and results}
\label{sec:theory}

\textbf{DFT simulations:} 

First-principles calculations were performed using density-functional theory (DFT) implemented in the SIESTA code using the PBE generalised-gradient approximation. A double-zeta polarised (DZP) basis set was used as generated by a 50 meV energy shift along with a split norm of  0.15, except for H, for which the splitnorm was of 0.5 [1,2]. Core electrons were replaced by norm-conserving Troullier-Martins pseudopotentials [5]  with cutoff radii of 1.54  Å for the s, p, d, and f channels of C; and 1.25 Å for H. The molecule was allowed to relax up to a force tolerance of 10 meV/Å.

\begin{figure*}[hbt]
\includegraphics[width=\columnwidth]{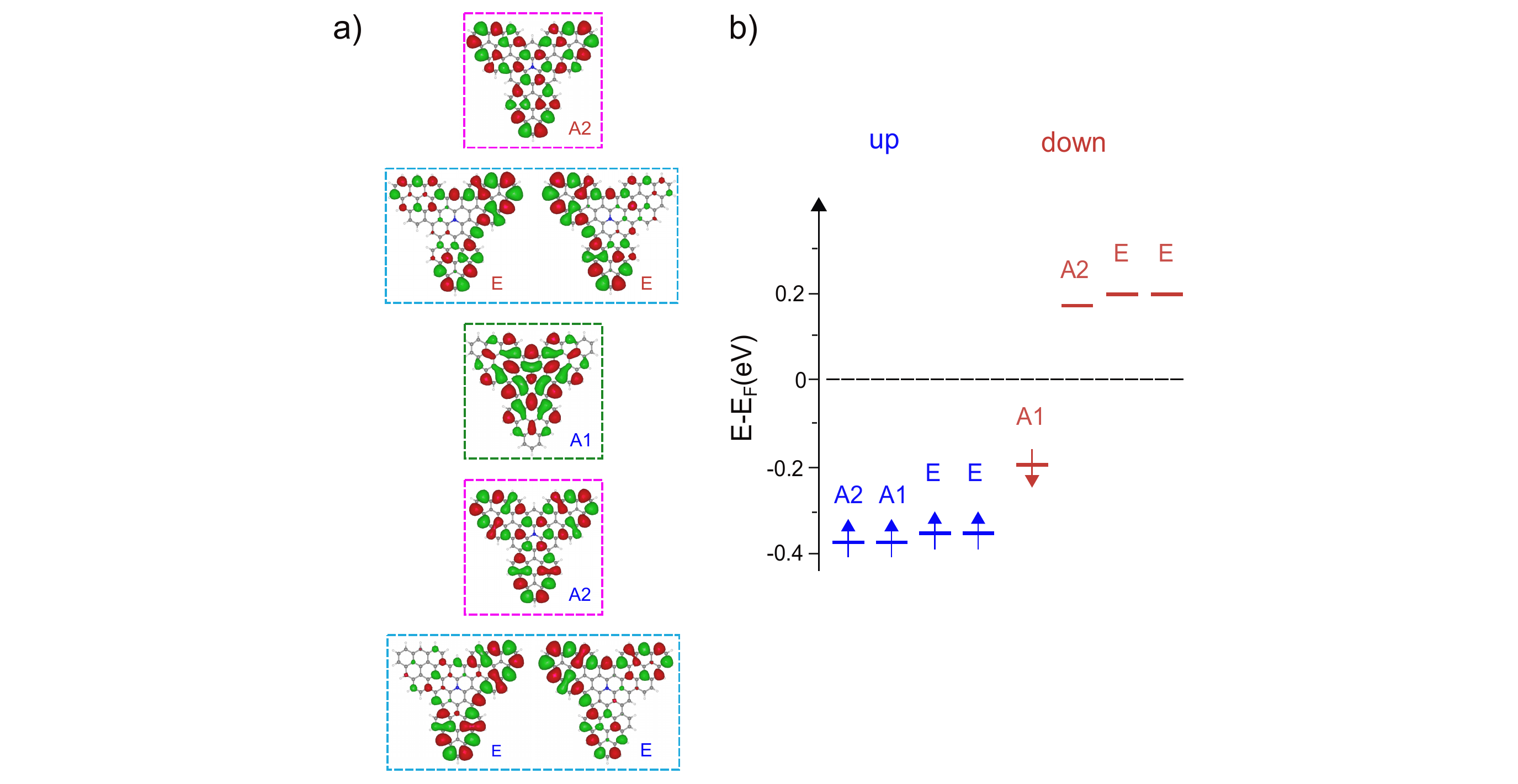}
\caption{a) Molecular orbital isosurfaces corresponding to the single-particle states shown in the energy level diagram in b) obtained from spin-polarized DFT calculations. The labels refer to the symmetry of each orbital, defined in terms of irreducible representations of the C$_{3v}$ symmetry point group. The singly occupied orbitals are three: two degenerate orbitals with E symmetry, and one with A2 symmetry. The only orbital with the highest intensity on the N site is the A1, which is fully occupied.}
 \label{figSI:DFT} 
\end{figure*}

\textbf{CASSCF calculations:} 

We performed CASSCF calculations on \textbf{TTAT} with the ORCA 5.0.2 package \cite{Neese2012}. First, the geometry was relaxed at the DFT level, with the PBE density functional and def2-SVP basis set. An auxiliary basis set was automatically generated with the AutoAux keyword. The resulting DFT orbitals and optimized geometry were then used as input for CASSCF, where the Complete Active Space consisted in 7 electrons fluctuating in 10 orbitals (CAS(7,10)). The optimized natural orbital with most fractional occupation had 1.82 electrons, while the natural orbital with less fractional occupation had 0.05 electrons. 

\textbf{CASCI calculations:} 


The molecular geometries were optimized in quartet states using density functional theory (DFT), as implemented in the FHI-AIMS software package \cite{Blum2009}, employing the PBE0 hybrid functional \cite{Adamo1999}. In these calculations, the Tkatchenko-Scheffler method was used to account for Van der Waals interactions. Given the open-shell and multi-radical nature of the molecules under investigation, the complete active space configuration interaction (CASCI) method was employed to obtain an accurate description of the wave function and electronic energies.
One- and two-electron integrals are constructed in the basis of molecular orbitals around the fermi energy  were derived using the quantum chemistry software ORCA \cite{Neese2012} using the orbitals from the restricted open-shell Hartree Fock (ROHF).

\begin{equation}
\label{integrals1body}
  t_{ij} =  \int \phi_i(\mathbf{r}) \left( -\frac{\hbar}{2m} \nabla^2 + V(\mathbf{r})
 \right) \phi_j(\mathbf{r}) d^3\mathbf{r}
\end{equation}
and
\begin{equation}
\label{integrals 2body}
 \mathcal{V}_{ijkl} = \dfrac{1}{4\pi \epsilon_0} \int \dfrac{\phi_i(\mathbf{r})\phi_{j}(\mathbf{r}^\prime)\phi_{k}(\mathbf{r}^\prime)\phi_{l}(\mathbf{r})}{\vert \mathbf{r} - \mathbf{r}^\prime   \vert}d^3\mathbf{r} d^3\mathbf{r}^\prime,
\end{equation}
where the indices $i,j,k,l$ denote molecular orbitals, extending across the set of orbitals chosen as the active space (CAS(11,11) for aza-triangulene and CAS(12,12 for full carbon analogue of aza-triangulene)). The one-electron potential $V(\mathbf{r})$ in Eq. \ref{integrals1body} encompasses the ionic potentials and the contributions from electrons in the occupied inactive orbitals, that is,
\begin{equation*}
    V(\mathbf{r}) 
     = 
     \dfrac{1}{4\pi \epsilon_0}
     \sum_\gamma 
     \frac{e Z_\gamma}{\vert \mathbf{R}_\gamma - \mathbf{r} \vert }
     + 
     \sum_\lambda \int 
     \dfrac{\vert \phi_\lambda(\mathbf{r}^\prime) \vert ^2 }{\vert \mathbf{r} - \mathbf{r}^\prime \vert}  
     d^3\mathbf{r},
\end{equation*}
where index $\gamma$ runs over the nuclei, $Z_\gamma$ is the charge of the nuclei and the index $\lambda$ runs over occupied inactive molecular orbitals given by $\phi_{\lambda}(\mathbf{r})$, which are always doubly-occupied in the possible Slater determinants of our calculation. Subsequently, we construct the many-body {\it ab initio} molecular Hamiltonian $\hat{\mathcal{H}}_{\text{CAS}}$ using these coefficients:

\begin{equation}
\label{molecular hamiltonian}
  \hat{\mathcal{H}}_{\text{CAS}} = \sum_{i,j, \sigma} t_{ij}\hat{c}^\dagger_{i \sigma}\hat{c}_{j \sigma} + \sum_{i,j,k,l, \sigma,\sigma^\prime} \mathcal{V}_{ijkl}\hat{c}^\dagger_{i \sigma}\hat{c}^\dagger_{j \sigma^\prime}\hat{c}_{k \sigma^\prime}\hat{c}_{l \sigma}.
\end{equation}

where $\hat{c}_{i\sigma}$ ($\hat{c}^\dagger_{i\sigma}$) denotes the annihilation (creation) operator of electron with spin $\sigma$ in $i$-th orbital. The full many-body Hamiltonian Eq. (\ref{molecular hamiltonian}) was diagonalized in our in-house code to obtain the many-body wave function $\Psi$ given by linear combination of Slater determinants.

\begin{figure*}[hbt]
\includegraphics[width=\columnwidth]{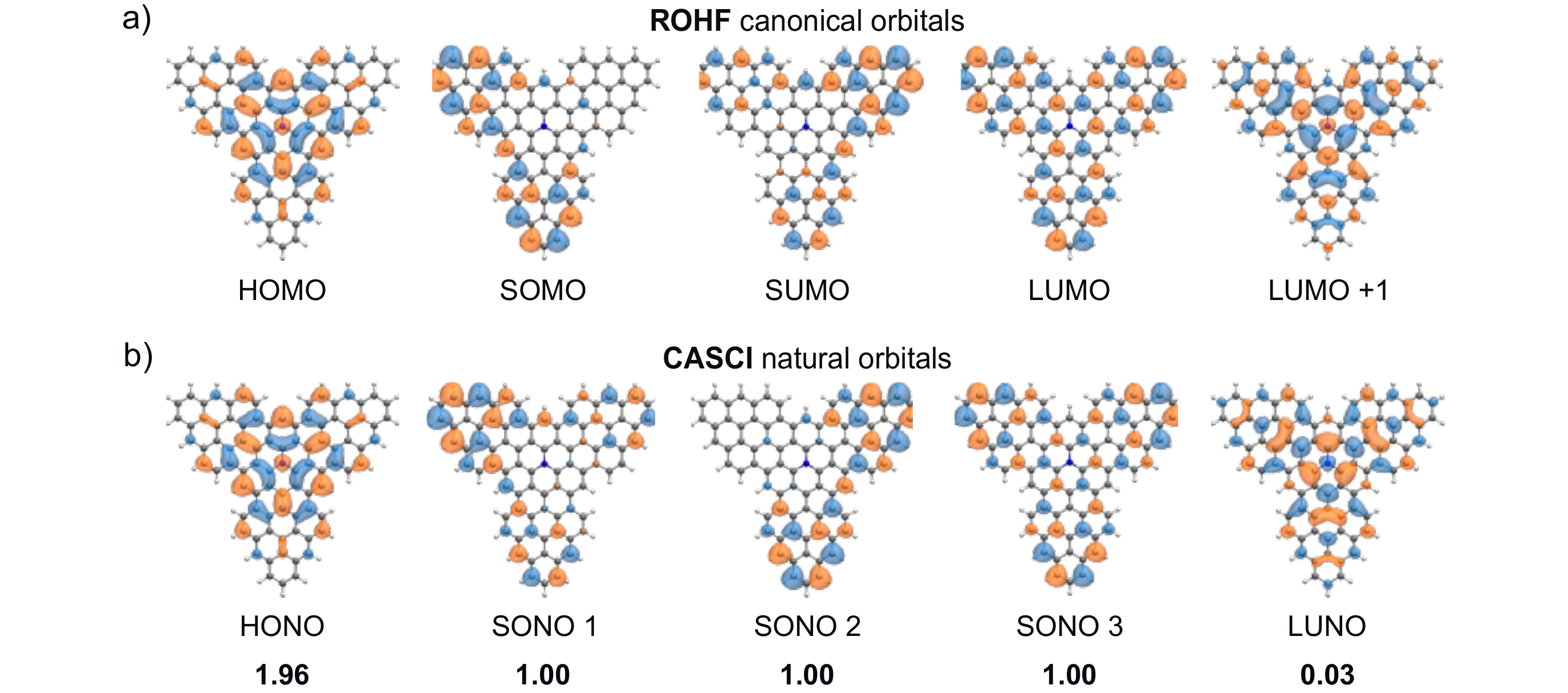}
\caption{a) Restricted open-shell Hartree-Fock (ROHF) orbitals and (b) complete active space configuration interaction (CASCI) natural orbitals of \textbf{TTAT}. The numbers below the natural orbitals indicate their fractional electronic occupation. As for the CASSCF calculations reported in the main text, three singly occupied natural orbitals (SONO) are obtained.}
 \label{figSI:FigS7} 
\end{figure*}

To get the number of unpaired electrons in the molecule, we have constructed the  one-particle density matrix, $\rho_{ij} = \langle \Psi | \hat{c}_i^\dagger \hat{c}_j | \Psi \rangle$ from the ground state many-body CASCI wavefunction. Natural orbitals are eigenvectors obtained from diagonalization of the one-particle reduced density matrix $\rho$, which positive eigenvalues  represent the occupations of the natural orbitals. The natural orbitals whose occupations have fractional values significantly different from integer values of 2 or 0 contribute to the number of unpaired electrons in the molecule. Fig.~\ref{figSI:FigS7}b shows the natural orbitals obtained for the aza-triangulene molecule.

\textbf{Natural Transition Orbitals (NTO):} 

\begin{figure*}[hbt]
\includegraphics[width=0.9\columnwidth]{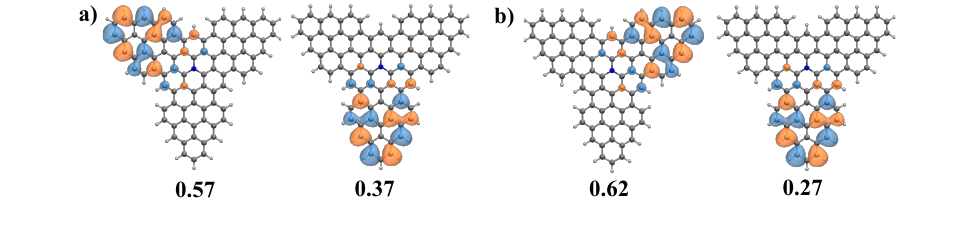}
\caption{NTOs for the spin excitation from quartet ground state to a) first doublet and b)  second doublet that is degenerate.}
\label{figSI:FigS8} 
\end{figure*}

To simulate the $dI/dV$ maps corresponding to IETS spin excitation maps, we have calculated the Natural Transition Orbitals (NTOs) \cite{Martin2003},  which correspond to the electronic transition density matrix of single spin flip process from the quartet ground state to the doublet excited states. NTO orbitals are obtained from the diagonalization of the matrix $TT^{\dagger}$, where the matrix $T$ is given by elements $$T_{jk}=\langle \Psi_{\text{doublet}} \vert \hat{c}^{\dagger}_{j \uparrow}\hat{c}_{k \downarrow}  \vert \Psi_{\text{quartet}}\rangle,$$ where the indices $j,k$ run over the orbitals of active space. We have constructed the $T_{tk}$ matrices from the many-body ground and first excited state CASCI wavefunctions and Fig.~\ref{figSI:FigS8} displays the calculated NTOs with corresponding amplitudes.

\textbf{Kondo Orbitals (KO):}

Kondo orbitals are calculated by diagonalizing the Hamiltonian derived from the multi-channel Anderson model, which considers the many-body multiplet structure of molecules obtained from the CASCI calculation for the neutral ground state and virtual charge states as described in the Ref. \cite{CalvoFernandez2024}. For the calculation of scattering amplitude corresponding to virtual processes that occur in the intermediate charge state with one more and one less electron of the molecular state, we have taken the five low-energy multiplets in charge states. The amplitude from the high-energy multiplets is negligible because the reduced Lehmann amplitude is divided by the energy difference between the neutral state and the energy of the charge multiplet. 

\begin{figure*}[hbt]
\includegraphics[width=0.9\columnwidth]{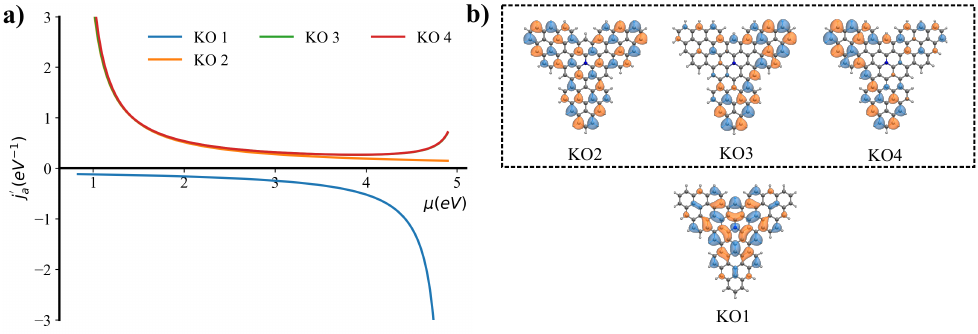}
\caption{Results from the multi-orbital Kondo analysis, used to simulate the Kondo spatial distribution map presented in the main text. (a) Coupling constants $j_{a}$ computed as a function of the chemical potential for each Kondo orbital (KO). (b) Orbital isosurfaces of the four KOs with non-zero coupling to the conduction electrons of the substrate at a chemical potential of 3 eV. Three of these orbitals (highlighted in the dashed box) correspond to channels with antiferromagnetic coupling ($j_{a}>0$) to the electron bath and therefore are involved in the many-body Kondo screening process.}
\label{figSI:FigS9} 
\end{figure*}

\textbf{Dyson Orbitals:} 
To accurately interpret differential conductance ($dI/dV$) maps for molecules exhibiting significant multireference character, it is necessary to move beyond the single-determinant molecular orbital framework typically provided by density functional theory (DFT) calculations.
In this context, we have constructed the Dyson orbitals to simulate the $dI/dV$ maps, as they are more relevant for the single-electron removal and addition processes in STM \cite{Zuzak2024, Ortiz2020a}. Dyson orbitals corresponding to the spatial negative ion resonance (NIR) and positive electron affinity (PEA) are constructed as
\begin{equation}
  \phi_{\text{NIR}}(\mathbf{r}) = \langle \Psi^{N - 1} |   \hat{c} (\mathbf{r}) | \Psi^N \rangle.
\end{equation}

\begin{equation}
  \phi_{\text{PEA}}(\mathbf{r}) = \langle \Psi^{N + 1} | \hat{c}^\dagger (\mathbf{r})  | \Psi^N \rangle.
\end{equation}
where $\Psi^{N}$, $\Psi^{N-1}$, and $\Psi^{N+1}$ are the many-body wavefunction obtained from the CASCI calculations for the N, N-1 and N+1 electrons respectively.

\begin{figure*}[hbt]
\centering
\includegraphics[width=0.8\columnwidth]{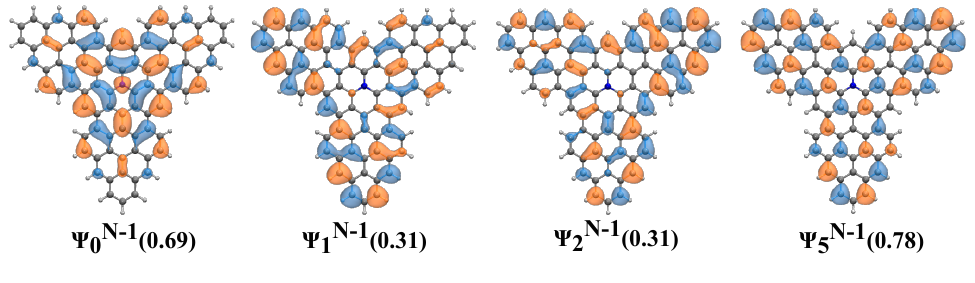}
\caption{Dyson orbitals for the process of removal of one electron with the norm of the wavefunction below.}
\label{figSI:FigS10} 
\end{figure*}

\begin{figure*}[hbt]
\centering
\includegraphics[width=0.7\columnwidth]{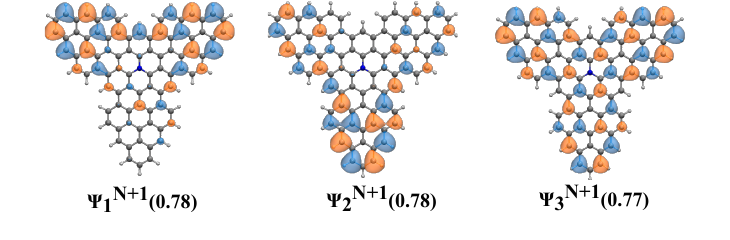}
\caption{Dyson orbitals for the process of adding one electron with the norm of the wavefunction below.}
\label{figSI:FigS11} 
\end{figure*}

Dyson orbitals provide a more comprehensive description of the electronic states involved in these tunneling events, capturing the electron removal/addition mechanisms during STS measurements. Fig.s~\ref{figSI:FigS10} and ~\ref{figSI:FigS11} displays calculated Dyson orbitals and their amplitudes for removal/addition of a single electron.

Theoretical $dI/dV$ maps of NTOs and Kondo orbitals were calculated by the Probe Particle Scanning Probe Microscopy (PP-SPM) code \cite{Krejci2017} for a CO-like tip. Heat maps are used to show the spatial current in the constant height mode for the theoretical dI/dV, where brighter colors represent the higher current values while black means no current. For NTOs  we have chosen the tip composed of pxpy (90\%) and s (10\%) orbitals while for Kondo orbitals the tip is composed of pxpy (85\%) and s (15\%) orbitals to match the experimental $dI/dV$ maps.

It is important to notice that, although these calculations do not directly include the surface, they consider variations of the charge state of the molecule, as discussed in the previous sections. Additionally, in our simulations, we assume that all the Kondo orbitals have identical Kondo coupling with the metal substrate, which is a good approximation since all the states are $\pi$ orbitals with similar localization.

\textbf{Fitting to a spin model:}

We performed the CASCI calculation in the basis of maximally localized orbitals (see Fig. 5a of the main text). In such representation, the wavefunctions of the ground and the excited states do not contain any fluctuations to states with doubly-occupied orbitals due to large Coulomb on-site interaction given by maximally localized character of orbitals.

Let us index with $j$ the three maximally localized orbitals. Then we construct the spin operators of the sites $j$: $\Vec{\hat{S}}_j = \sum_{\sigma \sigma^\prime} \Vec{P}_{\sigma \sigma^\prime}\hat{c}^\dagger_{j\sigma}\hat{c}_{j\sigma}, $ where $\Vec{P}_{\sigma \sigma^\prime}$ is the vector of components $\sigma, \sigma^\prime$ of the $\dfrac{1}{2}$ Pauli matrices.  We can then compute the spin correlations on this representation:
\begin{equation}
    \mathbb{C}_{j k} = \langle \Psi \vert \Vec{\hat{S}}_j \cdot \Vec{\hat{S}}_k  \vert \Psi   \rangle - \langle \Psi \vert \Vec{\hat{S}}_j \vert \Psi \rangle  \cdot  \langle \Psi \vert \Vec{\hat{S}}_k  \vert \Psi   \rangle,
\end{equation}
where $\Psi$ is the quartet ground state obtained from the CASCI calculation. 

Further, as seen in the many-body spectra, there are two degenerate duplets for the diagonalization of the hamiltonian in the sector $S_z= \dfrac{1}{2}.$

Therefore, our system is well described by the ferromagnetic Heisenberg model:
$$\hat{H} = J \left( \Vec{\hat{S}}_1 \cdot \Vec{\hat{S}}_2 + \Vec{\hat{S}}_1 \cdot \Vec{\hat{S}}_3 + \Vec{\hat{S}}_2 \cdot \Vec{\hat{S}}_3  \right),$$
where $J = 10 $ meV to reproduce the spin excitation and the homogeneous spin correlations. Notice that the identical coupling for all pairs of spins means that, in the duplet excited state, the trimer behaves as a frustrated magnet.

\textbf{Comparison with the all-carbon analogue:}

\begin{figure*}[hbt]
\centering
\includegraphics[width=0.6\columnwidth]{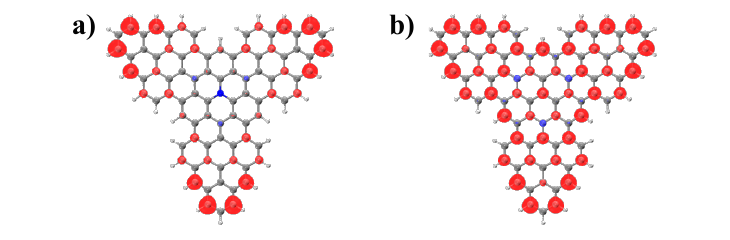}
\caption{Calculated spin density maps for the (a) aza-triangulene molecule obtained from CAS(11,11) and (b) fully carbon analogue obtained from CAS(12,12) calculations.}
\label{figSI:allcarbon1} 
\end{figure*}

In the case of the aza-triangulene, CAS(11,11) calculations determine three unpaired electrons located mainly on the three external triangulenes, where the spin density is located (Fig.~\ref{figSI:allcarbon1}a). In contrast, in the fully carbon analogue, CAS(12,12) calculations predict four unpaired electrons (as anticipated in the scheme in Fig.~1a in the main text) that are delocalized throughout the molecule (Fig.~\ref{figSI:allcarbon1}b): three unpaired electrons come from the three external triangulenes, and one unpaired electron is localized in the central part.

\begin{figure*}[hbt]
\centering
\includegraphics[width=\columnwidth]{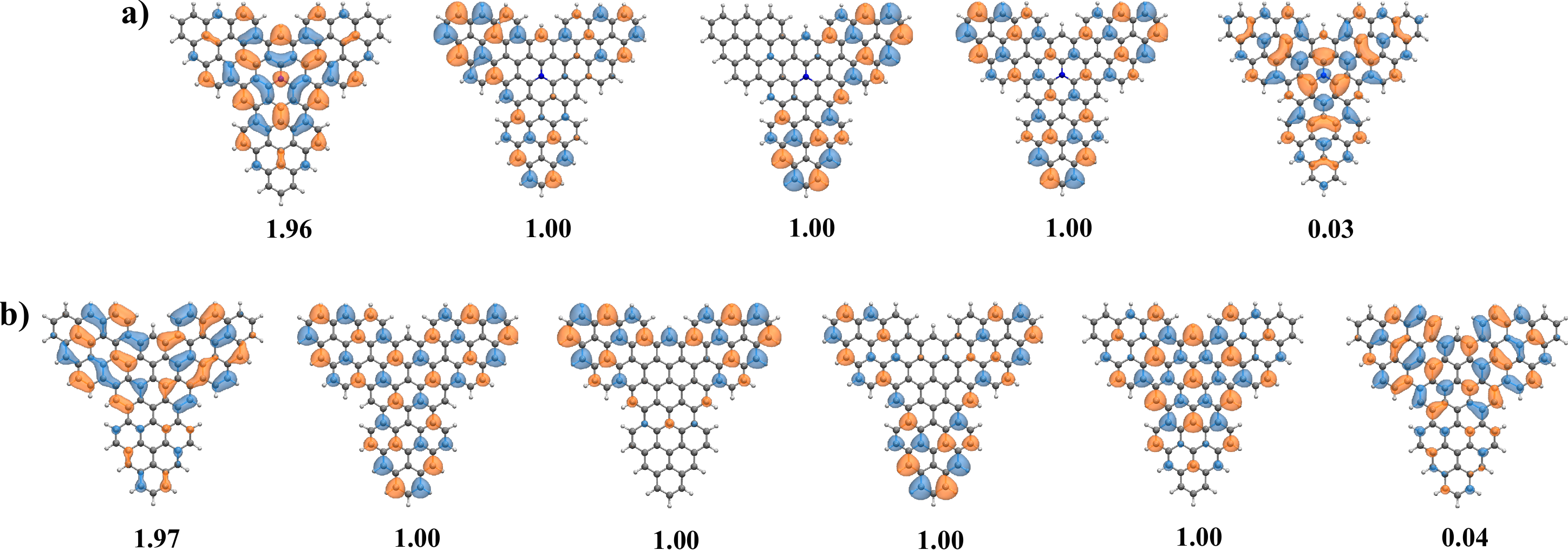}
\caption{Many-body natural orbitals (with the occupation indicated below), as obtained from the CAS(11,11) calculations for the (a) aza-triangulene molecule and (b) from CAS(12,12) calculations for the fully carbon analogue.}
\label{figSI:allcarbon2} 
\end{figure*}

From the occupation of the many-body natural orbitals obtained from CAS(11,11), it is clear that \textbf{TTAT} displays three orbitals with an occupation of 1, primarily located on the external triangulenes (Fig.~\ref{figSI:allcarbon2}a). Meanwhile, in the all-carbon analogue, there is an additional unpaired electron originating from the central part (Fig.~\ref{figSI:allcarbon2}b). The extra electron provided by the nitrogen atom in the N-substituted molecule quenches the central radical, which explains the absence of the central unpaired electron.

\begin{figure*}[h!]
\centering
\includegraphics[width=\columnwidth]{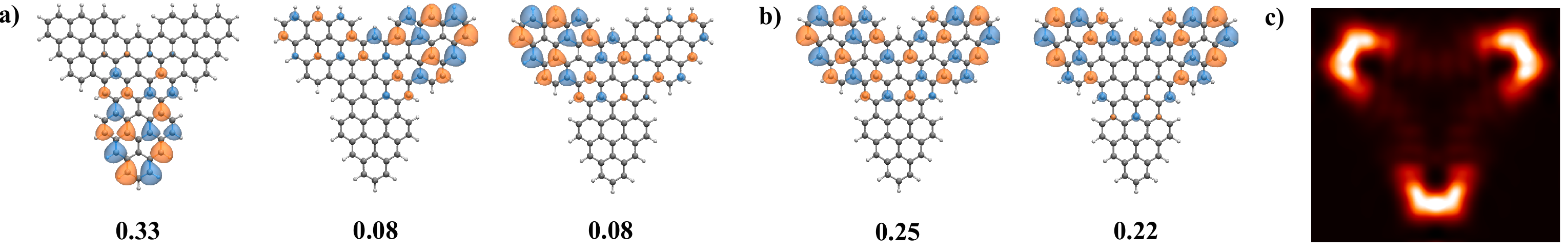}
\caption{Calculated NTOs from the many-body wavefunctions obtained from CAS(12,12) and their corresponding weights for the spin excitation from the quintet ground state to the degenerate (a) first triplet and (b) second triplet for the fully carbon molecule; c) simulated dI/dV map of NTOs of the spin excitation from the quintet ground state to the degenerate triplet excited states.}
\label{figSI:allcarbon3} 
\end{figure*}

In the purely carbon molecule, the ground state is a quintet ($S=2$), and the following excited states are two degenerate triplets with an energy gap of 242~meV from the quintet ground state. Fig.~\ref{figSI:allcarbon3} shows the dominant Natural Transition Orbitals of spin excitations from the ground state to the lowest excited states. The spin excitation from the quintet to the degenerate triplet states involves only the spins on the edges. 
Notably, it does not involve the central spin, further indicating that the exchange coupling between the edge spins differs from the coupling between the central spin and the edge spins.
The following two excited states are open-shell singlets, which are degenerate and positioned 590~meV higher in energy than the quintet ground state. These states involve spin flips of the unpaired electron at the central part of the carbon-based triangulene.

The exchange coupling (J) between the spins on the external triangulenes in the pure carbon molecule is symmetric, as evidenced by the fact that the first excited state is a doubly degenerate triplet. However, the coupling strength is an order of magnitude higher than that of the N-substituted molecule. Additionally, in the all-carbon molecule, the exchange coupling between the edge spins and the central spin is nearly twice as strong as the coupling between the edge spins, resulting in a double degeneracy in the open-shell singlet. 

\newpage
\setstretch{1.0}
\providecommand{\latin}[1]{#1}
\makeatletter
\providecommand{\doi}
  {\begingroup\let\do\@makeother\dospecials
  \catcode`\{=1 \catcode`\}=2 \doi@aux}
\providecommand{\doi@aux}[1]{\endgroup\texttt{#1}}
\makeatother
\providecommand*\mcitethebibliography{\thebibliography}
\csname @ifundefined\endcsname{endmcitethebibliography}  {\let\endmcitethebibliography\endthebibliography}{}